\documentclass[twocolumn,superscriptaddress,pra,showpacs,longbibliography]{revtex4-1}
\usepackage[colorlinks=true,urlcolor=blue,citecolor=blue,linkcolor=blue]{hyperref} 
\usepackage[T1]{fontenc}
\usepackage[utf8]{inputenc}
\usepackage{amssymb}
\usepackage{graphicx}
\usepackage{amsmath,color}
\usepackage{mathrsfs}
\usepackage{float}
\usepackage{indentfirst}
\usepackage{booktabs}

\makeatletter

\def\eqa{\begin{eqnarray}}
\def\eea{\end{eqnarray}}
\newcommand{\eq}{\begin{equation}}
\newcommand{\ee}{\end{equation}}

\newcommand{\Eq}[1]{Eq.~(\ref{#1})}

\newcommand{\upcite}[1]{\textsuperscript{\cite{#1}}}

\makeatother

\usepackage[english]{babel}

\begin{document}

\title{Ising Antiferromagnet in the 2D Hubbard Model with Mismatched Fermi Surfaces}

\author{Jan Gukelberger}
\affiliation{D{\'e}partement de Physique and Institut quantique, Universit{\'e} de Sherbrooke, Sherbrooke, Qu{\'e}bec, J1K 2R1, Canada}
\author{Lei Wang}
\affiliation{Beijing National Lab for Condensed Matter Physics and Institute
of Physics, Chinese Academy of Sciences, Beijing 100190, China}
\author{Lode Pollet}
\affiliation{Department of Physics and Arnold Sommerfeld Center for Theoretical Physics, Ludwig-Maximilians-Universit{\"a}t M{\"u}nchen, Theresienstrasse 37, 80333 Munich, Germany}

\begin{abstract}
We study the phase diagram of the two-dimensional repulsive Hubbard model with spin-dependent anisotropic hopping at half-filling. The system develops Ising antiferromagnetic long-range order already at infinitesimal repulsive interaction strength in the ground state. Outside the perturbative regime, unbiased predictions for the critical temperatures of the Ising antiferromagnet are made for representative interaction values by a variety of state-of-the-art quantum Monte Carlo methods, including the diagrammatic Monte Carlo, continuous-time determinantal  Monte Carlo and path-integral Monte Carlo methods. Our findings are relevant to ultracold atom experiments in the $p$-orbital or with spin-dependent optical lattices. 
\end{abstract}
\maketitle


\section{Introduction}


The Hubbard model plays an important role in condensed matter research combining a plethora of physical phenomena such as the Mott insulator transition, magnetism and (un)conventional superconductivity. Despite this richness, exact analytical solutions have only been found in a few special cases, including the one-dimensional model~\cite{essler2005}, the atomic, and the non-interacting limit. Furthermore, the Hubbard model is numerically tractable in infinite dimensions by dynamical mean-field theory~\cite{Anonymous:z_AfEOwS}, on bipartite lattices at half-filling by determinantal quantum Monte Carlo methods, and in the infinite-$U$ limit on ladder geometries with the density matrix renormalization group~\cite{Liu2011}. Going away from either of these special limits poses tremendous challenges to our theoretical understanding. Reference~\cite{LeBlanc:2015ha} summarizes the presently known results of the Hubbard model from a wide range of numerical algorithms. 

\begin{figure}[tbp]
\centering
\includegraphics[width=8cm]{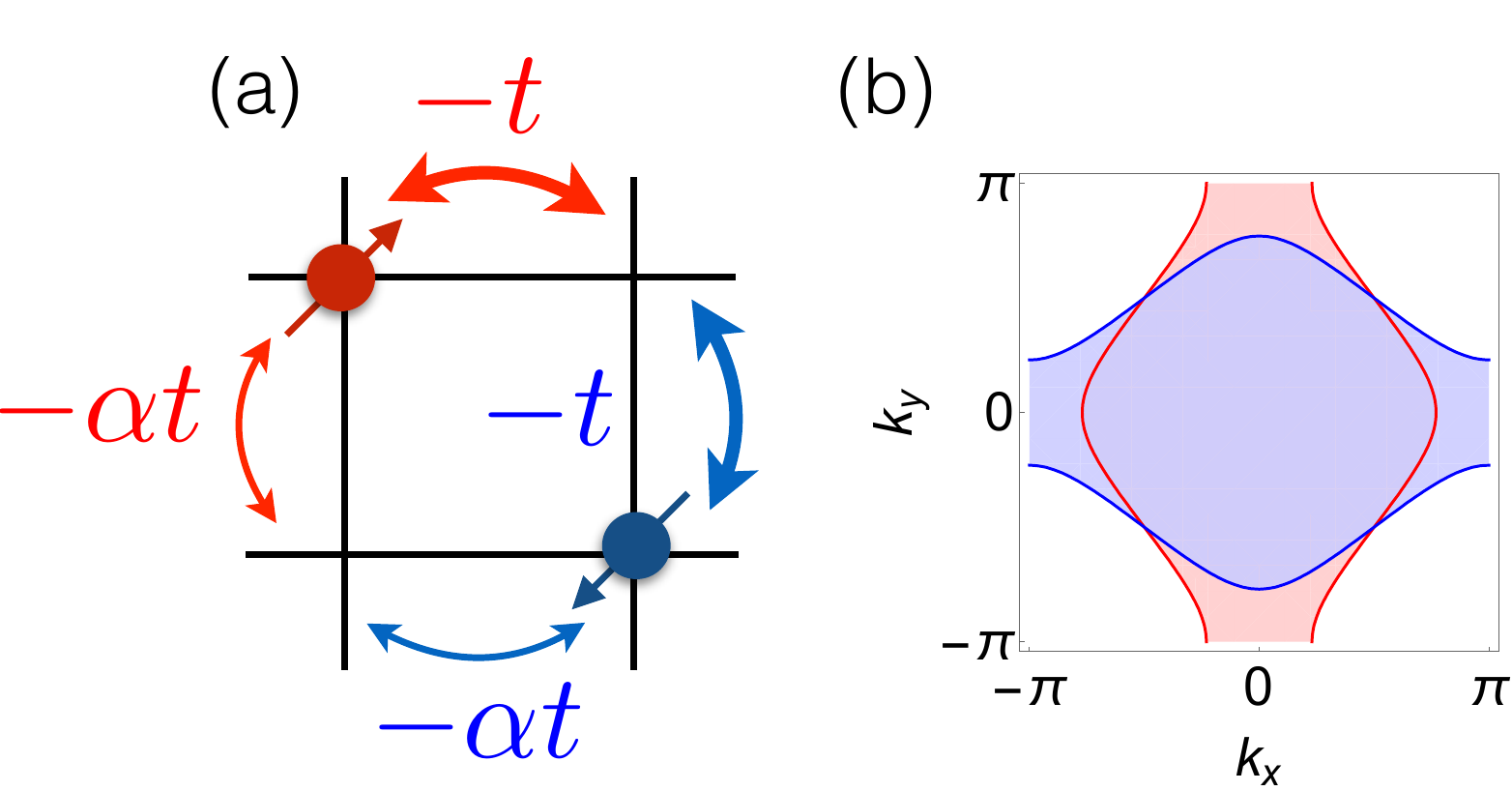}
\caption{(Color online) (a) Hopping amplitude of the model (\ref{eq:model}). (b) Mismatched Fermi surfaces of the two spin species shown for $\alpha = 0.75$.}
\label{fig:FS}
\end{figure}

In this paper, we add another parameter regime which can  be exactly solved numerically: we study the phase diagram of  the repulsive Hubbard model with spin-dependent anisotropic hopping by three different kinds of unbiased quantum Monte Carlo (QMC) algorithms. 
As we will see, some of these methods work only in certain parameter regimes but can be more efficient when they are applicable.
The Hamiltonian reads
\begin{eqnarray} 
\hat{H} = &-&\sum_{\sigma\in\{\uparrow,\downarrow\}}\sum_{{\pmb{\nu}\in\{\pmb{x,y}\}}} \sum_{\pmb{r}}\left(t_{\pmb{\nu}\sigma}\hat{c}^{\dagger}_{\pmb{r},\sigma} \hat{c}_{\pmb{r}+\pmb{\nu},\sigma} +h.c.\right) \nonumber \\ &+& U \sum_{\pmb{r}}\left(\hat{n}_{\pmb{r},\uparrow}-\frac{1}{2}\right)\left(\hat{n}_{\pmb{r},\downarrow }-\frac{1}{2}\right)
\label{eq:model}. 
\end{eqnarray}
Specifically, we consider the spin-dependent anisotropic hopping amplitude $t_{x\uparrow}=t_{y\downarrow}=t$ and $t_{x\downarrow}=t_{y\uparrow}=\alpha t$, where $\alpha \in [0,1]$ is a tuning parameter. The hopping is stronger along x(y) direction for spin up(down) fermions, shown in Fig.~\ref{fig:FS}(a). It leads to a spin-dependent nematic distortion of the Fermi surface in the reciprocal space shown in Fig.~\ref{fig:FS}(b). 
Physically, cold atomic systems~\cite{Bloch:2008gla, Esslinger:2010ex} may be well suited to study this system. The Hubbard model has been realized years ago~\cite{Jordens2008, Schneider2008} and with fermionic microscopes antiferromagnetic correlations have been measured~\cite{Boll2016, Parsons2016, Cheuk2016}, which can now extend over the entire system size and realize a Heisenberg antiferromagnet~\cite{Mazurenko2016}. 
The hopping anisotropy can either be realized with spin-dependent optical lattices, or due to the anisotropic shape of the Wannier function on the $p$-band of an optical lattice. The last term of (\ref{eq:model}) denotes an onsite repulsive interaction with $U>0$. We focus on magnetic order of the model (\ref{eq:model}) in the half-filled case on a square lattice.  
  
Diagonalizing the single-particle part of the Hamiltonian, one has dispersions $\epsilon_{k}^{\uparrow} = -2t \cos(k_{x}) -2\alpha t \cos(k_{y})$ and $\epsilon_{k}^{\downarrow} = -2 \alpha t \cos(k_{x}) -2 t \cos(k_{y})$. The noninteracting bandwidth is thus $W=4(1+\alpha)t$.
Spin nematicity explicitly breaks the spin SU(2) symmetry and removes the divergence in the density of states at the Fermi energy. However the perfect Fermi surface nesting with wave vector $\pmb{Q}=(\pi, \pi)$ is still preserved. As a consequence, the longitudinal spin susceptibility $\chi^{\sigma\sigma}(\pmb{Q})=\frac{1}{N}\sum_{k}\frac{f(\epsilon_{k}^{\sigma})-f(\epsilon^{\sigma}_{k+Q})}{\epsilon_{k}^{\sigma} - \epsilon_{k+Q}^{\sigma}}$ still diverges at zero temperature while the transverse spin susceptibility $\chi^{+-}(\pmb{Q})=\frac{1}{N}\sum_{k}\frac{f(\epsilon_{k}^{\uparrow})-f(\epsilon^{\downarrow}_{k+Q})}{\epsilon_{k}^{\uparrow} - \epsilon_{k+Q}^{\downarrow}}$ saturates to a constant value (Here, $f(\cdot)$ is the Fermi-Dirac distribution). Therefore, a weak-coupling analysis predicts Ising antiferromagnetic (AF) order already at infinitesimally strong repulsive interaction. 

The strong coupling limit of the model (\ref{eq:model}) was studied in the context of $p$-orbital Mott insulators~\cite{Wu:2008ea, Zhao:2008ev}. It reduces to a spin-$1/2$ XXZ model with Ising anisotropy, which favors an antiferromagnetic Ising ground state. At intermediate interaction strength, the system exhibits a crossover from a weak-coupling spin-density-wave state to the strong-coupling AF Ising state, similar to the case of  the 3D half-filled Hubbard model. However, since the Ising state breaks only a discrete $\mathbb{Z}_{2}$ symmetry, it has a finite critical temperature, even in two dimensions. 

The above considerations continue to hold in the fully anisotropic case $\alpha=0$. In this limit, the kinetic part is purely one-dimensional -- i.e. the motion of a spin is limited to a row or a column of the 2D lattice -- whereas the density-density interactions on each site connect the two spin species and make the system effectively two dimensional. Therefore, as we will show with unbiased worldline QMC simulations, the system still possesses a finite critical temperature.   

 
References~\cite{Feiguin:2009iu,Feiguin:2011jz,Gukelberger:2014dba} studied the model (\ref{eq:model}) in the doped attractive case in search of an elusive Bose liquid  and exotic superfluid states. It turns out that close to half filling the most stable phase is an incommensurate density wave state, which is related to the AF Ising state of the repulsive model discussed above upon a particle-hole transformation. It was also remarked in Ref.~\cite{Gukelberger:2014dba} that in the fully anisotropic limit the particle number of each column and row  is separately conserved. This allows one to perform unbiased worldline QMC calculations by mapping the fermions to quantum spins, similar to what was done in~\cite{Xu:2015gx}, and which served as a benchmarking tool for the diagrammatic Monte Carlo calculations.

The model (\ref{eq:model}) is different from the one studied in Ref.~\cite{Berg:2012ie}, which studied onset of antiferromagnetism in a similar Fermi surface geometry. However, Ref.~\cite{Berg:2012ie} doubles the fermion species to avoid the fermion sign problem. In addition, the current study focuses on the half-filled case where the AF Ising state is strongly enhanced due to the commensurate filling. 

The organization of the paper is as follows.  In Sec.~\ref{sec:methods} we summarize the Monte Carlo methods used in this paper and comment on their advantages and disadvantages. In Sec.~\ref{sec:results} we report on results obtained by various QMC calculations, where Sec.~\ref{sec:fullyanisotropic} contains results on the fully anisotropic case of the model (\ref{eq:model}) and Sec.~\ref{sec:generalanisotropic}  results for general anisotropic cases. Section~\ref{sec:summary} summarizes our main findings and discusses their implications for future experimental and theoretical studies.

\section{Methods \label{sec:methods}}
In this section we summarize the three different quantum Monte Carlo techniques used to study the model \Eq{eq:model}: Path Integral Monte Carlo simulations with worm-type updates (Worm), diagrammatic Monte Carlo simulations (DiagMC), and continuous-time determinantal Monte Carlo simulations (LCT-QMC). 
Table~\ref{tab:methods} summarizes their main features and allows one to quickly read off the method of choice. In their domain of applicability all three methods yield unbiased results on the physical observables. Whenever there is an overlap in their application range we have checked that they give consistent results. In the subsections below we explain in more detail the specifics of all three methods for the anisotropic Hubbard model.

\begin{table}
\caption{\label{tab:methods} A comparison of the QMC methods used. For the Worm and LCT-QMC methods the sign-positive regimes are mentioned in the table. They scale linearly and cubically in the system volume, respectively, and both linearly with the inverse temperature. DiagMC simulations work directly in the thermodynamic limit. In practice, open boundary conditions are used in the Worm simulations.}
\begin{tabular}{lccc}
\toprule
Method          & Anisotropy & Filling  & Interaction \\
\midrule 
Worm~\upcite{Prokofev1998,Pollet2007}       & $\alpha=0$ & arbitrary  & arbitrary \\
DiagMC~\upcite{VanHoucke2010, gukelberger2015diss}          & arbitrary & arbitrary   & $U \lesssim 4 t$ \\
LCT-QMC~\upcite{Iazzi:2015hi, 2015PhRvB..91w5151W}     & arbitrary & half filling &    arbitrary \\
 \bottomrule 
 \end{tabular}
\end{table}

\subsection{Path-integral Monte Carlo (Worm)~\label{sec:wormalgorithm}}
In the fully anisotropic limit, which is where the Worm algorithm can be applied, the model \Eq{eq:model} reduces to
\begin{eqnarray}
H_{\alpha=0} & = &  -t \sum_{\mathbf{r}} \hat{c}^{\dagger}_{\mathbf{r}, \uparrow} c_{\mathbf{r} + \mathbf{x}, \uparrow} - t \sum_{\mathbf{r}} \hat{c}^{\dagger}_{\mathbf{r}, \downarrow} c_{\mathbf{r} + \mathbf{y} , \downarrow} + \textrm{h.c.} \nonumber \\
{} &{} & + U \sum_{\mathbf{r}} \left(\hat{n}_{\mathbf{r}, \uparrow} - \frac{1}{2}  \right) \left( \hat{n}_{\mathbf{r}, \downarrow} - \frac{1}{2}   \right).
\label{eq:alpha0}
\end{eqnarray}
The hopping is one-dimensional, implying that for each row (column) the number of up (down) particles is conserved.  By translational invariance we expect that all or none of these symmetries are simultaneously broken. 
As a consequence of the 1D character, individual rows and columns can be mapped onto hard-core bosons at any density through the celebrated Jordan-Wigner transformation~\cite{Cazalilla2011, Giamarchi_book}, which in turn allows us to use path-integral Monte Carlo simulations with worm-type updates~\cite{Prokofev1998}, here in the implementation of Ref.~\cite{Pollet2007}.
Spin densities and density-density correlations functions, which we measure in order to identify the phase transition, are not affected by the Jordan-Wigner transformation and identical for the original fermions and the simulated hard-core bosons.
For ease of the Jordan-Wigner transformation, we use open boundary conditions. This comes at the price of greater finite size effects through the influence of the boundary terms, which is however minor in light of the mapping to a positive expansion for all filling factors and the linear scaling of the Worm algorithm with system size and inverse temperature. For $\alpha \neq 0$ the Worm algorithm has a sign problem leading to an exponential scaling in the system volume and inverse temperature.

\subsection{Diagrammatic Monte Carlo (DiagMC)}

The diagrammatic Monte Carlo (DiagMC) method evaluates Feynman diagrammatic expansions by means of a stochastic process that samples sums over diagram topologies and internal variables on equal grounds~\cite{prokofev2007bdm, prokofev2008fpp}.
Our implementation for the Hubbard model~\cite{Gukelberger:2014dba,gukelberger2015diss}, which is based on diagrams with bare propagators $G_0$ and interactions $U$, is not directly applicable within a magnetically ordered phase. Therefore, we detect a continuous phase transition to AF order by monitoring the divergence of the magnetic susceptibility on approaching the critical temperature. This aspect is different from the other two Monte Carlo methods, which are not formulated in the thermodynamic limit.
To this end we sample the self-energy $\Sigma_\sigma(k)$ and the irreducible scattering vertex in the particle-hole channel $\Gamma^{ph}_{\sigma \sigma'}(Q, k, k')$ for fixed total four-momentum $Q=(\pmb{Q}, i \Omega_m=0)$ with $\pmb{Q}=(\pi, \pi)$ the AF ordering vector.
According to the Bethe-Salpeter equation
\begin{align}
\chi(Q) &= \frac{\chi^{0}(Q)}{1 + \chi^{0}(Q) \Gamma(Q)},
\end{align}
the susceptibility $\chi(Q)$ diverges when the largest eigenvalue of the kernel $-\chi^{0}(Q) \Gamma(Q)$ reaches unity.
The above should be read as a matrix equation for the generalized susceptibility $\chi^{\sigma \sigma'}(Q; k, k')$ in spin and four-momentum space. Furthermore, the particle-hole bubble $\chi^{0}_{\sigma \sigma'}(Q; k, k') = G_\sigma(k+Q/2) G_\sigma(k-Q/2) \delta_{\sigma, \sigma'} \delta(k-k')$ is the diagonal product of two one-particle propagators and the one-particle propagators in turn are calculated from the self-energy via Dyson's equation.

With DiagMC the system is directly simulated in the thermodynamic limit, but the diagrammatic series for the irreducible quantities $\Sigma$ and $\Gamma$ must be restricted to orders $n \leq N_*$ because the sign of a fermionic series vanishes factorially with diagram order $n$.
All DiagMC results must therefore be extrapolated in the cutoff parameter $N_* \to \infty$. The uncertainty in this extrapolation is typically the dominant contribution to the error bars and the extrapolation may be impossible when the series does not converge quickly enough. This happens frequently if the interaction is too strong, e.g.\ $U \gtrsim W$. 
For models like the half-filled Hubbard model where determinantal QMC methods do not suffer from the sign problem, the sign-problem-free method will generally yield smaller error bars than DiagMC under comparable computational efforts.
The main advantage of DiagMC is that it can equally well be applied away from half filling, where simulations with other QMC methods are often unfeasible due to a severe sign problem.
Additionally, the comparison of finite-size extrapolations (e.g.\ from path-integral or determinantal QMC) with finite-order extrapolations from DiagMC yields a very nontrivial crosscheck that all systematic errors in the different methods are under control.

\subsection{Continuous-time determinantal Monte Carlo (LCT-QMC)}

We employ the continuous-time quantum Monte Carlo method  scaling linearly in $\beta$ (LCT-QMC)~\cite{Iazzi:2015hi, 2015PhRvB..91w5151W} to study the model (\ref{eq:model}) at general anisotropies on finite lattices. The LCT-QMC methods perform continuous-time interaction expansion of the partition function and evaluate each expansion as a matrix determinant.  Thanks to recent progress on the fermion sign problem~\cite{Huffman:2014fj, Li:2015jf, Wang:2015hm, PhysRevLett.116.250601} these matrix determinants can be shown to be nonnegative. There is no sign problem in the simulation despite the mismatched Fermi surfaces: The crucial conditions are half filling and the presence of bipartite lattices. The implementation of the LCT-QMC simulation is similar to the recent study of the mass-imbalanced Hubbard model~\cite{Liu:2015kx}. As the signature of the phase transition we measure the staggered magnetization square according to the Wick's theorem in the LCT-QMC simulations.  

Compared to the path-integral Monte Carlo method of Sec.~\ref{sec:wormalgorithm}, the drawback of the  LCT-QMC algorithm is that it scales cubically with the system size. We are therefore limited to system sizes $L\le 24$ for the LCT-QMC results. The advantage, however, is that one is able to study also finite anisotropy ratios and systems with periodic boundary conditions can be simulated without further constraints because the method does not rely on the Jordan-Wigner mapping.

\section{Results\label{sec:results}}

In this section we first present our results for the fully anisotropic case, followed by the results for the more general case.
The unit of energy is set by the hopping $t=1$ unless explicitly noted otherwise.

\subsection{The fully anisotropic model \label{sec:fullyanisotropic}}

Below we use bosonization arguments to get an intuitive and analytical understanding of  the phase diagram at zero temperature, followed by quantum Monte Carlo simulations addressing the phase transition at finite temperature. We will see that the ground state is always gapped and ordered in spin space, whereas at finite temperature a  $\mathbb{Z}_{2}$ transition between a normal liquid and an antiferromagnet is found. Unless otherwise specified, we limit ourselves to the half-filled case.

\subsubsection{Bosonization considerations of the ground state}

Thanks to the one-dimensional nature of the hopping, each row and column can be bosonized separately. Following the notation and the formulas of App. D in the standard book (Ref.~\cite{Giamarchi_book}) we write the harmonic action for  row $j$ as
\begin{equation}
H_{\uparrow}^j = \frac{1}{2 \pi} \int dx \, u_{\uparrow}^j K_{\uparrow}^j ( \nabla \theta^j_{\uparrow}(x))^2 + \frac{u_{\uparrow}^j}{K_{\uparrow}^j}( \nabla \phi^j_{\uparrow}(x))^2,
\end{equation}
where $u_{\uparrow}$ is a velocity and $K_{\uparrow} $ the dimensionless Luttinger parameter. The fields $\nabla \phi$ and $\nabla \theta$ are proportional to the sum and the difference of right and left movers, respectively. For a column $\bar{j}$ a similar expression can be written down with the replacements ${\uparrow} \leftrightarrow {\downarrow}$, $x \leftrightarrow y$ and $j \leftrightarrow \bar{j}$. We still need to investigate the Hubbard term, which couples the spin densities on intersecting rows and columns, and take care of the filling factor. The density in bosonized form is
\begin{equation}
\rho_{\uparrow}^j(x) = \rho_0 - \frac{1}{\pi} \nabla \phi^j(x) + \rho_0 \sum_{p \neq 0} e^{i 2 p (\pi \rho_0 x - \phi^j_{\uparrow}(x))},
\end{equation}
with $\rho_0 = 1/2$ at half filling. 

Introducing the charge $\phi^{j \bar{k}}_{\rho} = (\phi^{j}_{\uparrow} + \phi^{\bar{k}}_{\downarrow} )/\sqrt{2}$ and the spin $\phi^{j \bar{k} }_{\sigma} = (\phi^{j}_{\uparrow} - \phi^{\bar{k}}_{\downarrow} )/\sqrt{2}$ fields we get a non-oscillating term $\cos(\sqrt{8} \phi^{j \bar{k}}_{\rho})$ resulting from the Hubbard interaction, as well as a term $\cos(\sqrt{8} \phi^{j \bar{k}}_{\sigma})$.
If we assume that translational invariance is not broken, then the fields for all $j$ and $\bar{j}$ are the same, and the cosines become relevant in both sectors; {\it i.e.}, similar to the 1D Hubbard model with spin and repulsive interactions at half filling  the charge sector is always massive at zero temperature. Its gap can be exponentially small $\ln \Delta \sim -1/\sqrt{U}$ in the weak-coupling regime (cf. Eq.~\eqref{eq:Tc} below). However, in contrast to the 1D Hubbard model, the spin sector cannot remain a spin liquid because of the 2D nature of the lattice (which we see in the bosonization via the presence of the second cosine term). The system therefore orders into an Ising antiferromagnet in order to lower its energy.
Away from half filling, similar arguments can be applied leading to incommensurate spin density waves, in line with the weak-coupling and DiagMC results of Ref.~\cite{Gukelberger:2014dba} for the attractive case.

\subsubsection{Monte Carlo results for the Ising transition at finite temperature}

\begin{figure}[t]
\includegraphics[angle=-90,width=8cm]{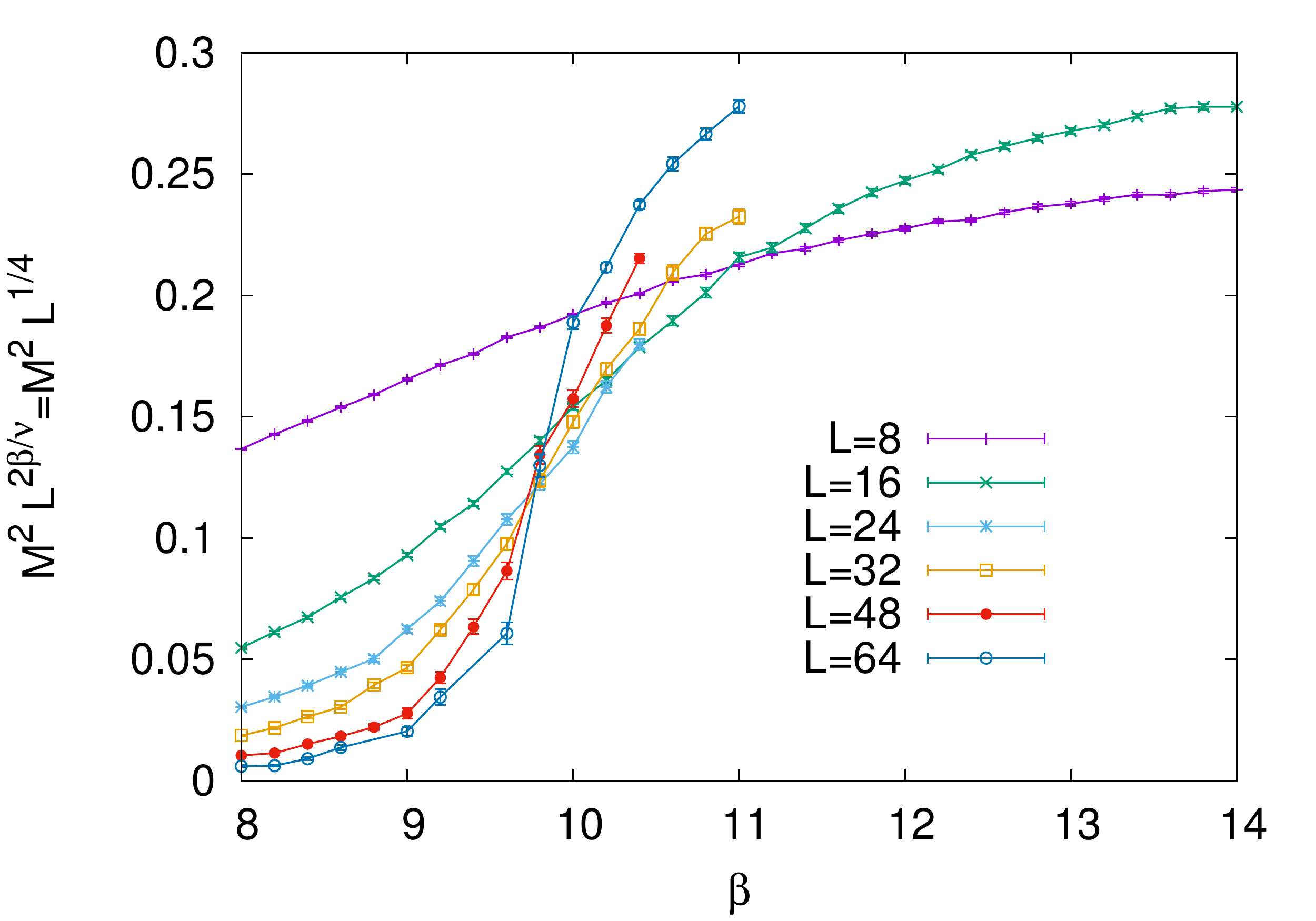}
\caption{(Color online) Finite size scaling of the staggered magnetization squared as a function of the inverse temperature $\beta$ for different system sizes of linear length $L$ for a fully anisotropic Hubbard model with $U=3$. Error bars for the biggest system sizes could be larger than shown (see text). Nevertheless, the critical temperature can be estimated as $\beta_c = 9.6(4)$ justifying the 2D Ising universality class.} 
\label{fig:U3}
\end{figure}

At finite temperature one expects a phase transition between a normal liquid and an Ising antiferromagnet with critical exponents belonging to the 2D classical Ising universality class. 
In order to test this, we performed large scale Monte Carlo simulations using the Worm algorithm and computed the expectation value square $\langle M_{\rm st}^2 \rangle$ and fourth power $\langle M_{\rm st}^4 \rangle$ of the staggered magnetization $M_{\rm st}$,
\begin{equation}
M_{\rm st} = \sum_{\mathbf{r}=(x,y)} (-1)^{x+y}  ( \hat{c}^{\dagger}_{\mathbf{r}, \uparrow}  \hat{c}_{\mathbf{r} , \uparrow} - \hat{c}^{\dagger}_{\mathbf{r}, \downarrow}  \hat{c}_{\mathbf{r} , \downarrow} ).
\end{equation}
Finite size scaling theory predicts, in leading order, that the curves $\langle M_{\rm st}^2 \rangle L^{2 \beta/\nu}$ intersect in a single point. Here, $L$ is the linear system size of the system, $\beta$ is the critical exponent for the order parameter which is $\beta=1/8$ for the 2D classical Ising model, and $\nu=1$ is the critical exponent for the correlation length. This is shown in Fig.~\ref{fig:U3}, where we see that the system sizes $L=8$ and $L=16$ are too small to be taken into account in the finite size analysis. For system sizes $L=24$ and larger we get curves that intersect, within error bars, in almost a single point when the staggered magnetization squared is multiplied with the correct power of the system size, $L^{1/4}$, in agreement with the critical exponents of the classical 2D Ising class. 

Although the linear scaling in the system volume of the Worm algorithm suggests it should be the method of choice in the absence of a sign problem, the Worm algorithm is nevertheless not well equipped to study this Ising transition because the worms are confined to single rows and single columns. 
The spin-resolved single-particle density matrix is hence one-dimensional and decays exponentially in the gapped phase: 
The algorithm is in the spin sector not better than a single spin-flip algorithm for a classical 2D Ising model. We have checked for $\beta = 8$ that  the integrated autocorrelation time increases linearly with $L$  with  a very large prefactor. Very close to the transition point, additional critical slowing down takes place with a dynamical exponent $z \approx 2$, just as in the single spin-flip algorithm for a classical 2D Ising model. 
To give an idea, for $L=32, \beta = 9.6$, we find a value around 100 with a binning analysis, where each measured value taken into account in the binning analysis is already an average of 1000 Monte Carlo measurements. Measurements were taken after 1000 Monte Carlo worm updates to compensate for the size of the system. The total calculation lasted several CPU-months per data point and resulted in more than half a million measurements, but that is barely enough.
An immediate consequence is that the fluctuations on the Binder cumulant are an order of magnitude worse than the ones in Fig.~\ref{fig:U3}, and are therefore less precise to locate the phase transition.
We have also successfully repeated this analysis for $U=4$ (not shown) with $\beta_c = 6.2(5)$. We leave for future work whether a new algorithm can be devised which combines a spin-cluster algorithm with the Worm algorithm in order to overcome this critical slowing down.
Despite the present algorithm's inefficiencies we have obtained results for larger systems than accessible with any other method. The finite size scaling is further validated by DiagMC results for the thermodynamic limit in Sec.~\ref{sec:diagmc-results}.

We also tried a similar analysis for the ground state assuming the universality class of the 3D classical Ising spin model (not shown). For system sizes up to $L=128$ we failed to find a single crossing point: curves for the staggered magnetization squared multiplied with $L^{2 \beta/\nu}$ have all the same shape with the steep part shifting parallel to lower values of $U$ with increasing $L$. This is consistent with the ground state being ordered for any $U$, in line with the bosonization arguments. Since the charge gap opens exponentially slowly for low values of $U$ there is of course no chance of observing the ground state in a brute-force numerical approach in the small $U$ limit.

\subsection{General anisotropic case~\label{sec:generalanisotropic}}

At general anisotropy $\alpha>0$ both fermion species can hop in the 2D plane. Therefore an effective bosonic description can no longer hold. In the following we show that the system has a weak-coupling instability to antiferromagnetic order for all values of $\alpha$. Then we obtain unbiased results for the transition temperature at intermediate interaction using two fermionic QMC methods (LCT-QMC and DiagMC) and cross-check the results. In general one anticipates that in the highly anisotropic case ($\alpha \ll 1$) the critical temperature approaches the one determined by the bosonic Worm calculation in Sec~\ref{sec:fullyanisotropic}, whereas the critical temperature drops to zero when $\alpha$ approaches unity, restoring the full $SU(2)$ rotational symmetry. 

\subsubsection{Weak coupling analysis} \label{sec:weakcoupling}

Particle-hole symmetry of the half-filled model \eqref{eq:model} ensures that the Fermi surfaces for the two spin species individually are nested with respect to the AF wave vector $\pmb{Q}=(\pi, \pi)$ independent of the anisotropy $\alpha$. Therefore the longitudinal spin susceptibilities always have a logarithmic divergence
\begin{align} \label{eq:chilong}
\chi^{0}_{\uparrow \uparrow}(\pmb{Q}) &= \chi^{0}_{\downarrow \downarrow}(\pmb{Q}) \sim -\ln \frac{T}{T_F}
\end{align}
for $T \to 0$. Here $T_F$ denotes the Fermi temperature.
In contrast, nesting between $\uparrow$- and $\downarrow$-Fermi surfaces is destroyed by an anisotropy $\alpha \neq 1$, so that the transverse spin susceptibilities $\chi^{0}_{+-}$, $\chi^{0}_{-+}$ saturate to finite values at low temperature.
Since there are no first-order pairing instabilities in the particle-particle channel for repulsive interactions, longitudinal (Ising) antiferromagnetism is the only instability to leading order in $U$.

In the first-order approximation to the Bethe-Salpeter kernel in the longitudinal particle-hole channel the irreducible vertex is replaced by the bare interaction
\begin{align}
\Gamma(Q)_{\sigma,k;\sigma',k'} = U \delta_{\sigma, -\sigma'} + \mathcal{O}(U^2) .
\end{align}
This yields an eigenvalue
\begin{align}
\lambda &= U \chi^{0}_{\sigma \sigma}(\pmb{Q}),
\end{align}
which grows logarithmically according to \eqref{eq:chilong} and will hence reach unity for arbitrarily small $U$ at a critical temperature 
\begin{align}
T_c &= T_F \exp(-c/U),
\label{eq:Tc}
\end{align}
which has the typical form of a BCS-type weak-coupling instability ($c$ is the constant prefactor of the logarithmic divergence in \eqref{eq:chilong}).

In summary, a weak-coupling analysis predicts a general low-temperature instability of the Fermi liquid towards Ising-type antiferromagnetic order (cf. the previous section) --- except at the isotropic point $\alpha=1$ where longitudinal and transverse channels become degenerate and magnetic order at finite temperature is ruled out by the continuous spin rotation symmetry (as long as the system remains purely two-dimensional).
For weak coupling we expect the $T_c$ suppression to be confined to a very small region around the isotropic point because at exponentially low temperatures the physics is extremely sensitive to small Fermi surface mismatches.
Away from half filling and at $\alpha > 0$ the perfect nesting and hence the weak-coupling instability in the particle-hole channel is lifted. Then only second-order instabilities in the particle-particle pairing channel remain, leading to $p$-wave superfluidity in direct correspondence to the attractive case \cite{Gukelberger:2014dba}.

\subsubsection{DiagMC results} \label{sec:diagmc-results}

\begin{figure}[!htb]
\includegraphics[width=8cm]{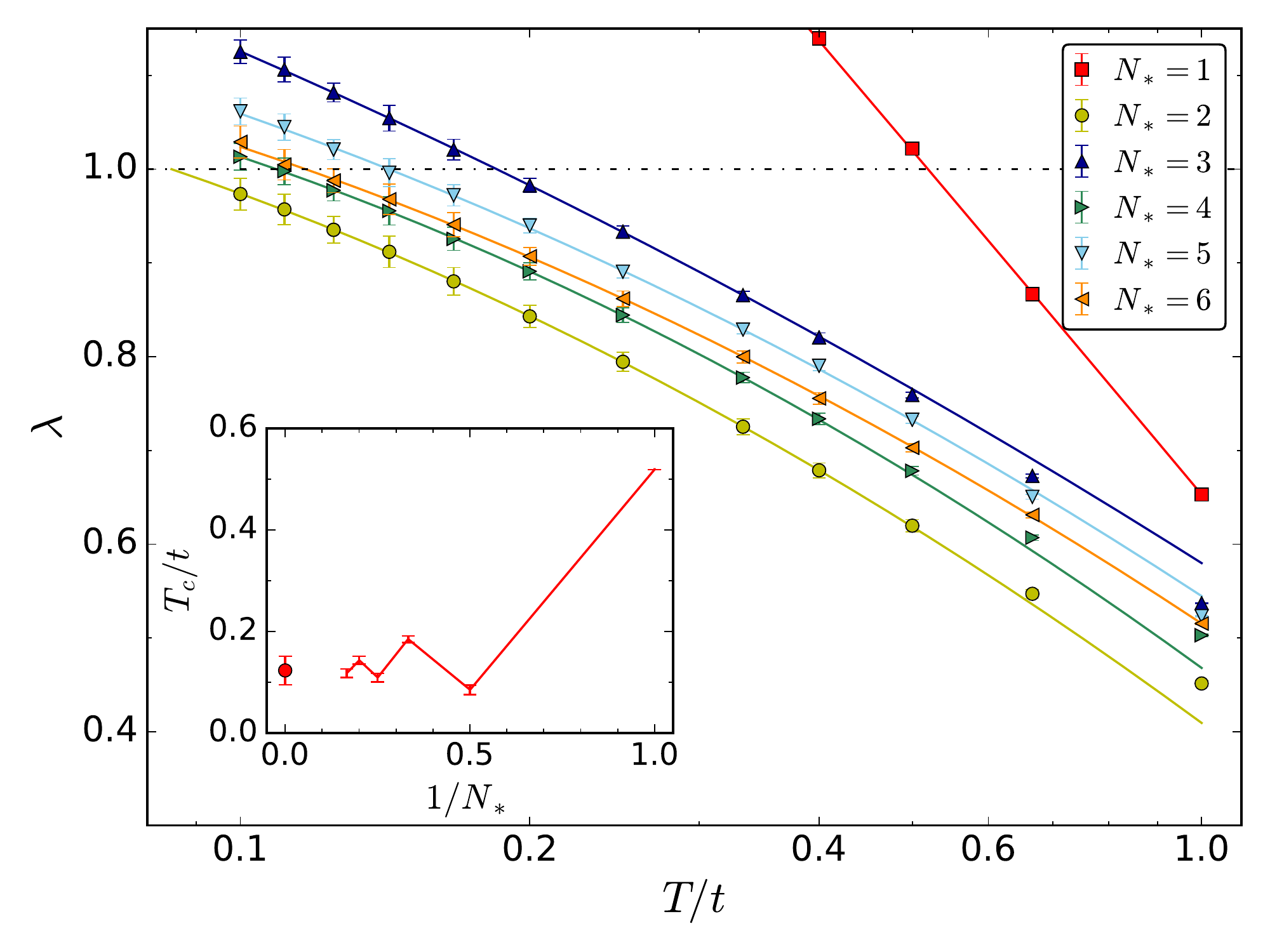}
\caption{(Color online) Leading Bethe-Salpeter eigenvalue $\lambda$ versus temperature $T$ from DiagMC simulations for $U=3$, $\alpha=0$ with cutoff order $N_* = 1, \dots, 6$. Lines are quadratic fits in $\log T$ used to interpolate the data around $T_c$. \textit{Inset:} Estimates of the transition temperature $T_c$ determined from these fits. The circle represents our extrapolation $T_c(N_* \to \infty) = 0.12(3) t$.} 
\label{fig:diagmc-alpha0}
\end{figure}

\begin{figure}[!htb]
\includegraphics[width=8cm]{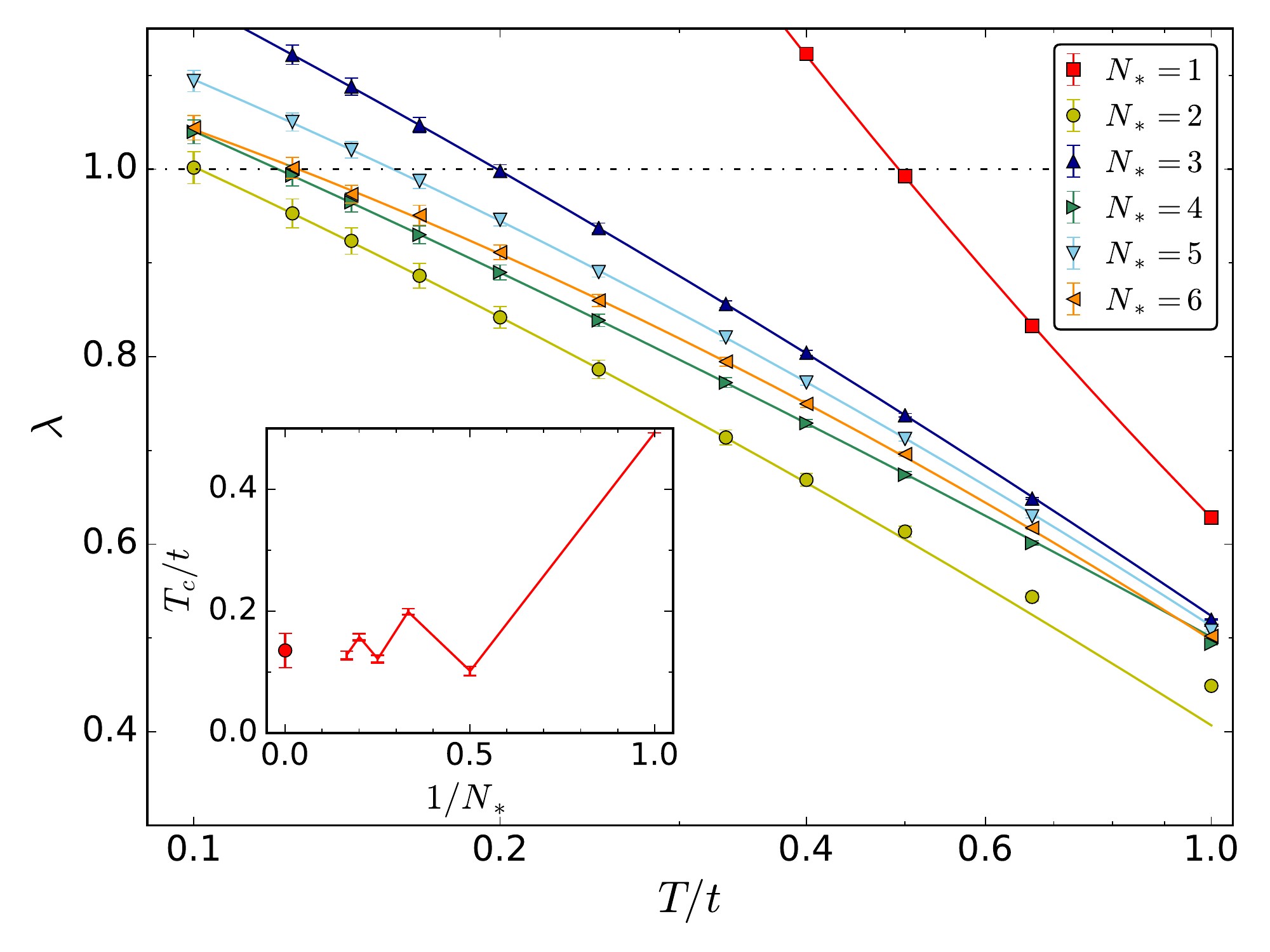}
\caption{(Color online) Like Fig.~\ref{fig:diagmc-alpha0}, but for  for $U=3$, $\alpha=0.75$. The extrapolated transition temperature is $T_c(N_* \to \infty) = 0.14(3) t$.} 
\label{fig:diagmc-alpha075}
\end{figure}

In order to go beyond the weak-coupling analysis we turn to DiagMC simulations, which can address arbitrary anisotropy. As shown in Figs.~\ref{fig:diagmc-alpha0} and \ref{fig:diagmc-alpha075} we track the leading Bethe-Salpeter eigenvalue for antiferromagnetic order. As the temperature is lowered, the eigenvalue grows and eventually crosses unity, causing a divergence of the AF susceptibility. While a cutoff order $N_*=1$ corresponds to a mean-field treatment and strongly overestimates the transition temperature, the eigenvalues for higher cutoffs converge reasonably quickly with a decaying even-odd oscillation: The eigenvalue for each order lies between the values from the next two smaller orders. We take the average of the three largest orders $N_*=4, 5, 6$ as extrapolation to infinite order and give error bars that cover these three finite-order results.
For the fully anisotropic model (Fig.~\ref{fig:diagmc-alpha0}) we obtain the transition point $\beta_c = 8.2 \pm 1.7$. This is consistent with the Worm result presented above. As expected, the DiagMC error bar is markedly larger than the one obtained with sign-problem-free bosonic QMC.
The results at general anisotropy $\alpha=0.75$ (Fig.~\ref{fig:diagmc-alpha075}) are very similar to the fully anisotropic case even though the kinetic terms are changed from one-dimensional to two-dimensional. In this case we obtain a slightly larger transition temperature $\beta_c = 7.5 \pm 1.4$.

\subsubsection{LCT-QMC results}

\begin{figure}[!b]
\includegraphics[width=8cm]{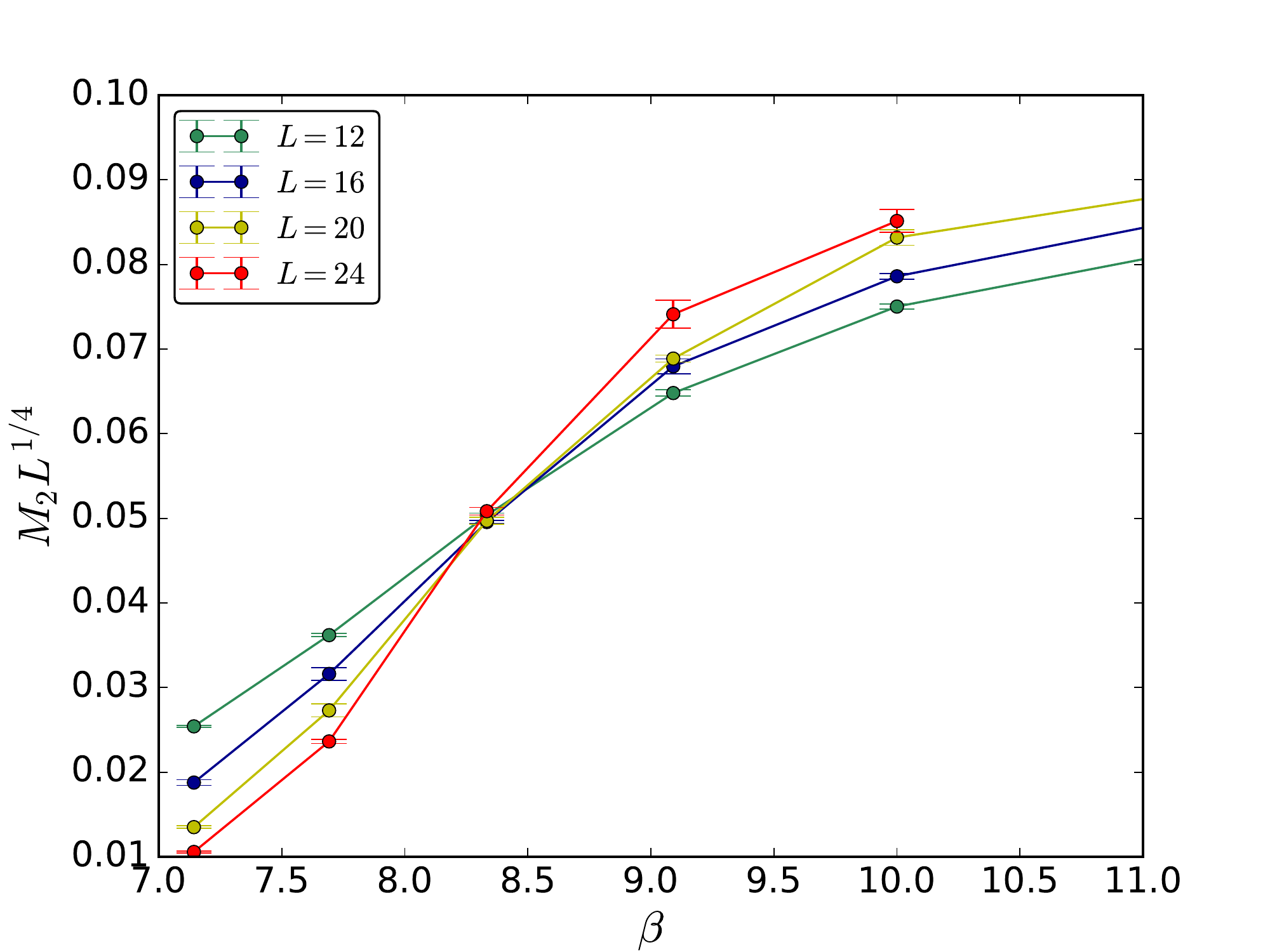}
\caption{(Color online)  Scaled staggered magnetization squared (according to the 2D Ising critical exponent) as a function of the inverse temperature $\beta$ (in units of the hopping amplitude $t$) for different system sizes of linear length $L$ for an anisotropic Hubbard model with $U/t=3,\alpha=0.75$. The critical temperature is estimated from the intersections to be $\beta_c t =8.5\pm0.5$.} 
\label{fig:LCT-QMC}
\end{figure}

\begin{figure}[!b]
\includegraphics[width=8cm]{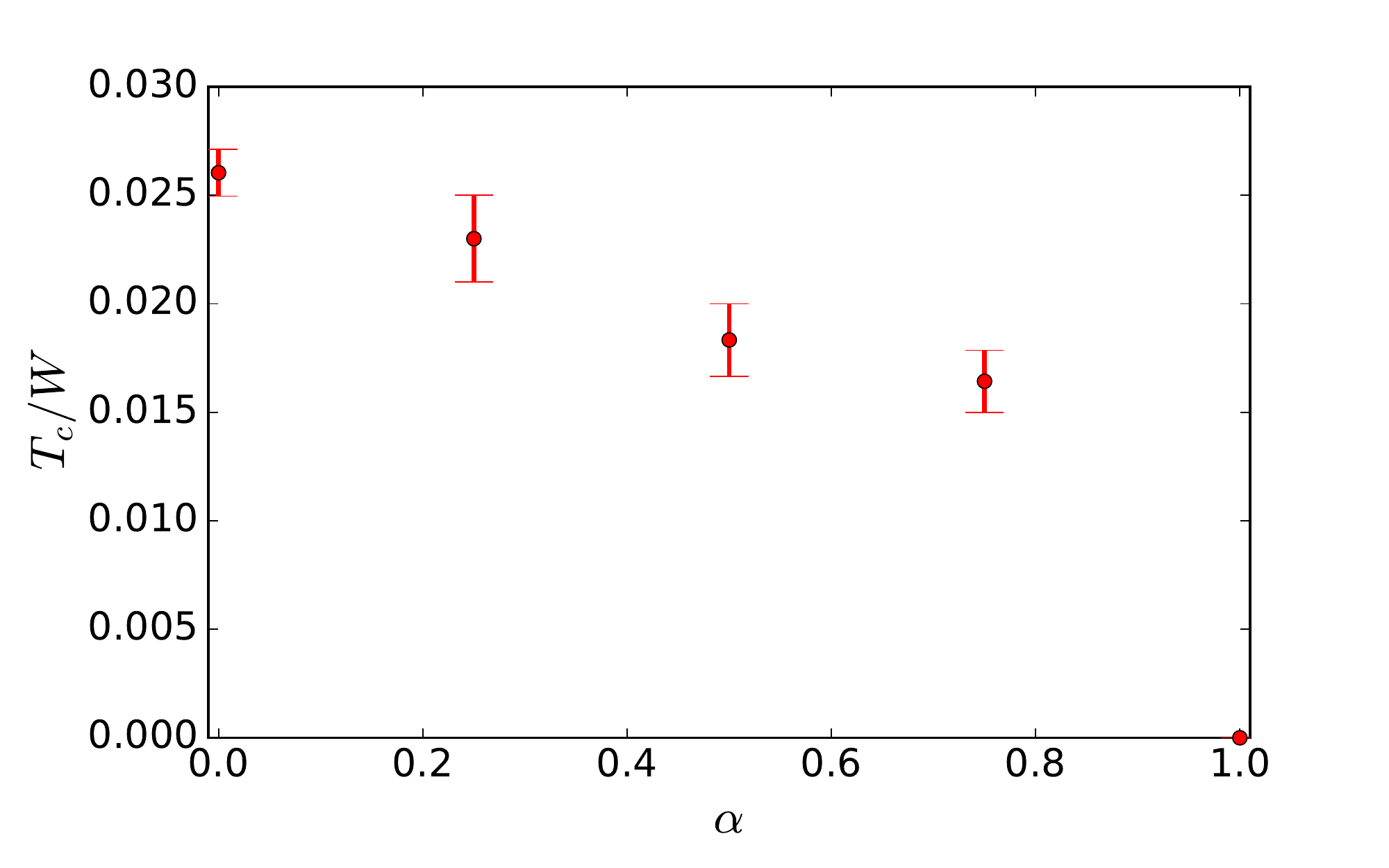}
\caption{(Color online) Critical temperature of the model (\ref{eq:model}) versus anisotropy at $U/t=3$. The critical temperature is measured in the unit of the noninteracting bandwidth $W=4(1+\alpha)t$. The data point at $\alpha=0$ is from the worm calculation (Fig.~\ref{fig:U3}) and the  critical temperature is known to be zero at $\alpha=1$~\cite{MerminWagnerTheorem1966}, while the other data are from the LCT-QMC calculation.} 
\label{fig:phasediag}
\end{figure}

Using the LCT-QMC method, we obtained the critical temperature at $U/t=3,\alpha=0.75$ by scaling the staggered magnetization according to the 2D Ising critical exponent as is shown in Fig.~\ref{fig:LCT-QMC}. The estimate is again in agreement with the critical temperature obtained by the DiagMC calculations. 
Figure~\ref{fig:phasediag} summarizes the critical temperature computed at different anisotropic ratios. The critical temperature measured in the unit of the bandwidth $W$ remains high from extreme ($\alpha\sim0$) to intermediate ($\alpha\sim0.75$) anisotropy. When measured in the unit of the hopping amplitude $t$ the transition temperature even rises with decreasing anisotropy. However, since $T_c$ should drop to zero in the isotropic case, it suggests a quite abrupt change of the critical temperature in the neighborhood of the isotropic point $\alpha\sim1$. This behavior is reminiscent of the XXZ model~\cite{1402-4896-32-4-016}, which applies in the strong coupling limit, and also appears natural in the weak-coupling limit (cf.\ Sec.~\ref{sec:weakcoupling} above). 

\section{Summary and experimental realization \label{sec:summary}}
Breaking of the discrete $\mathbb{Z}_{2}$ spin inversion symmetry results in an antiferromagnetic Ising state with finite critical temperature on a two-dimensional lattice. We presented quantitative predictions for the onset of antiferromagnetic Ising order in a Hubbard model with mismatched Fermi surfaces. Since the model (\ref{eq:model}) can be implemented using spin-dependent optical lattices or higher orbitals, testing these predictions would be a step stone to further approach the more exotic quantum phases at different filling and interaction~\cite{Gukelberger:2014dba} and on different lattice geometries~\cite{Wu:2008ea, Zhao:2008ev}.  

On the methodological side, we have shown how specific limits of the Hubbard model  with mismatched Fermi surfaces on two-dimensional lattices can be brought under full numerical control by using three different quantum Monte Carlo methods. These limits are (i) fully anisotropic spin-dependent hopping at any density with the Worm algorithm, (ii) anisotropic spin-dependent hopping at half filling with LCT-QMC, and (iii) arbitrary anisotropy and density at sufficiently weak interactions using DiagMC. The results of the methods are consistent with each other  within their domain of applicability and furthermore supported by analytical weak-coupling arguments. 

The main physical result is that we find a discrete $\mathbb{Z}_{2}$ symmetry breaking at a finite critical temperature towards an Ising antiferromagnet in the half-filled model. In view of our numerical results at intermediate interactions as well as the situation in the weak- and strong-coupling limits, a small anisotropy seems to be generically sufficient to create a large critical temperature.
One should notice that breaking the spin $SU(2)$ symmetry is crucial to obtain an Ising antiferromagnet with finite critical temperature in 2D. The strictly symmetrical distortion of the spin up and down Fermi surfaces shown in Fig.~\ref{fig:FS} is however not crucial. 
Away from half filling the weak-coupling limit predicts incommensurate spin density waves at extreme anisotropy $\alpha=0$ and $p$-wave superfluidity at general anisotropy $0 < \alpha < 1$. At finite interactions we expect the $\mathcal{O}(U)$ spin density wave instability to dominate over the $\mathcal{O}(U^2)$ superfluid instabilities in a finite region of the phase diagram at strong anisotropy and around half filling, similar to the situation in the attractive case~\footnote{Cf.\ Ref.~\cite{Gukelberger:2014dba}. Note that, while there is a direct mapping between results for the half-filled attractive and repulsive models, the doped repulsive model maps into the attractive model in a magnetic field and vice versa~\cite{ho2009qsh}.}.

Experimentally, the model (\ref{eq:model}) can be implemented using spin-dependent optical lattices~\cite{PhysRevLett.91.010407} or using higher orbitals~\cite{PhysRevLett.99.200405, Wirth:2010bh}. In both cases, tuning the  anisotropic hopping amplitudes of two species of atoms differently will lead to mismatched Fermi surfaces like those shown in Fig.~\ref{fig:FS}(b). Moreover, by using magnetic gradient modulation, Ref.~\cite{PhysRevLett.115.073002} has created a continuously tunable state-dependent optical lattice. 
In the extreme anisotropic and large $U$ limit the model (\ref{eq:alpha0}) approaches the 2D Ising model. The critical entropy per particle is $0.30647k_{B}$~\cite{PhysRev.65.117}, which is likely to be within reach in current experiment~\cite{Mazurenko2016}. The presence of an additional trapping potential is likely to reduce the required entropy~\cite{PhysRevLett.104.180401} because the metallic wings with incommensurate filling have higher entropy than the trap center. 

Detection of the phase transition would be easiest via a spin-resolved density measurement. Forming of the antiferromagnetic pattern would also leave a signature in spin-resolved time-of-flight images. Testing our predictions would be an important step towards approaching the more exotic quantum phases at different filling and interaction strengths~\cite{Gukelberger:2014dba} and on different lattice geometries~\cite{Wu:2008ea, Zhao:2008ev}, which are at the limits of numerical control.  

\section{Acknowledgment}
We thank Y.-H. Liu for helpful discussions and Wei Tang for providing the critical entropy value of the 2D Ising model. JG is supported by the Swiss National Science Foundation, LW  by the Ministry of Science and Technology of China under the Grant No.2016YFA0302400 and the start-up grant of IOP-CAS, and LP  by FP7/ERC starting grant No. 306897. The DiagMC calculations were run on the Mammouth cluster of Universit\'e de Sherbrooke, provided by the Canadian Foundation for Innovation, the Minist\`ere de l'\'Education des Loisirs et du Sport (Qu\'ebec), Calcul Qu\'ebec, and Compute Canada. The LCT-QMC calculations were run on the Tianhe-2 cluster of the National Supercomputer Center in Guangzhou. Simulations and data evaluation made use of the ALPS libraries~\cite{alps20,alps13}.

\bibliography{refs}{}

\begin{thebibliography}{46}%
\makeatletter
\providecommand \@ifxundefined [1]{%
 \@ifx{#1\undefined}
}%
\providecommand \@ifnum [1]{%
 \ifnum #1\expandafter \@firstoftwo
 \else \expandafter \@secondoftwo
 \fi
}%
\providecommand \@ifx [1]{%
 \ifx #1\expandafter \@firstoftwo
 \else \expandafter \@secondoftwo
 \fi
}%
\providecommand \natexlab [1]{#1}%
\providecommand \enquote  [1]{``#1''}%
\providecommand \bibnamefont  [1]{#1}%
\providecommand \bibfnamefont [1]{#1}%
\providecommand \citenamefont [1]{#1}%
\providecommand \href@noop [0]{\@secondoftwo}%
\providecommand \href [0]{\begingroup \@sanitize@url \@href}%
\providecommand \@href[1]{\@@startlink{#1}\@@href}%
\providecommand \@@href[1]{\endgroup#1\@@endlink}%
\providecommand \@sanitize@url [0]{\catcode `\\12\catcode `\$12\catcode
  `\&12\catcode `\#12\catcode `\^12\catcode `\_12\catcode `\%12\relax}%
\providecommand \@@startlink[1]{}%
\providecommand \@@endlink[0]{}%
\providecommand \url  [0]{\begingroup\@sanitize@url \@url }%
\providecommand \@url [1]{\endgroup\@href {#1}{\urlprefix }}%
\providecommand \urlprefix  [0]{URL }%
\providecommand \Eprint [0]{\href }%
\providecommand \doibase [0]{http://dx.doi.org/}%
\providecommand \selectlanguage [0]{\@gobble}%
\providecommand \bibinfo  [0]{\@secondoftwo}%
\providecommand \bibfield  [0]{\@secondoftwo}%
\providecommand \translation [1]{[#1]}%
\providecommand \BibitemOpen [0]{}%
\providecommand \bibitemStop [0]{}%
\providecommand \bibitemNoStop [0]{.\EOS\space}%
\providecommand \EOS [0]{\spacefactor3000\relax}%
\providecommand \BibitemShut  [1]{\csname bibitem#1\endcsname}%
\let\auto@bib@innerbib\@empty
\bibitem [{\citenamefont {Essler}\ \emph {et~al.}(2005)\citenamefont {Essler},
  \citenamefont {Frahm}, \citenamefont {G{\" o}hmann}, \citenamefont
  {Klümper},\ and\ \citenamefont {Korepin}}]{essler2005}%
  \BibitemOpen
  \bibfield  {author} {\bibinfo {author} {\bibfnamefont {Fabian H.~L.}\
  \bibnamefont {Essler}}, \bibinfo {author} {\bibfnamefont {Holger}\
  \bibnamefont {Frahm}}, \bibinfo {author} {\bibfnamefont {Frank}\ \bibnamefont
  {G{\" o}hmann}}, \bibinfo {author} {\bibfnamefont {Andreas}\ \bibnamefont
  {Klümper}}, \ and\ \bibinfo {author} {\bibfnamefont {Vladimir~E.}\
  \bibnamefont {Korepin}},\ }\href {\doibase 10.1017/CBO9780511534843} {\emph
  {\bibinfo {title} {The One-Dimensional Hubbard Model:}}}\ (\bibinfo
  {publisher} {Cambridge University Press},\ \bibinfo {address} {Cambridge},\
  \bibinfo {year} {2005})\BibitemShut {NoStop}%
\bibitem [{\citenamefont {Georges}\ \emph {et~al.}(1996)\citenamefont
  {Georges}, \citenamefont {Kotliar}, \citenamefont {Krauth},\ and\
  \citenamefont {Rozenberg}}]{Anonymous:z_AfEOwS}%
  \BibitemOpen
  \bibfield  {author} {\bibinfo {author} {\bibfnamefont {Antoine}\ \bibnamefont
  {Georges}}, \bibinfo {author} {\bibfnamefont {Gabriel}\ \bibnamefont
  {Kotliar}}, \bibinfo {author} {\bibfnamefont {Werner}\ \bibnamefont
  {Krauth}}, \ and\ \bibinfo {author} {\bibfnamefont {Marcelo~J}\ \bibnamefont
  {Rozenberg}},\ }\bibfield  {title} {\enquote {\bibinfo {title} {{Dynamical
  mean-field theory of strongly correlated fermion systems and the limit of
  infinite dimensions}},}\ }\href
  {http://rmp.aps.org/abstract/RMP/v68/i1/p13_1} {\bibfield  {journal}
  {\bibinfo  {journal} {Reviews of Modern Physics}\ }\textbf {\bibinfo {volume}
  {68}},\ \bibinfo {pages} {13} (\bibinfo {year} {1996})}\BibitemShut {NoStop}%
\bibitem [{\citenamefont {Liu}\ \emph {et~al.}(2012)\citenamefont {Liu},
  \citenamefont {Yao}, \citenamefont {Berg}, \citenamefont {White},\ and\
  \citenamefont {Kivelson}}]{Liu2011}%
  \BibitemOpen
  \bibfield  {author} {\bibinfo {author} {\bibfnamefont {Li}~\bibnamefont
  {Liu}}, \bibinfo {author} {\bibfnamefont {Hong}\ \bibnamefont {Yao}},
  \bibinfo {author} {\bibfnamefont {Erez}\ \bibnamefont {Berg}}, \bibinfo
  {author} {\bibfnamefont {Steven~R.}\ \bibnamefont {White}}, \ and\ \bibinfo
  {author} {\bibfnamefont {Steven~A.}\ \bibnamefont {Kivelson}},\ }\bibfield
  {title} {\enquote {\bibinfo {title} {Phases of the infinite $u$ hubbard model
  on square lattices},}\ }\href {\doibase 10.1103/PhysRevLett.108.126406}
  {\bibfield  {journal} {\bibinfo  {journal} {Phys. Rev. Lett.}\ }\textbf
  {\bibinfo {volume} {108}},\ \bibinfo {pages} {126406} (\bibinfo {year}
  {2012})}\BibitemShut {NoStop}%
\bibitem [{\citenamefont {LeBlanc}\ \emph {et~al.}(2015)\citenamefont
  {LeBlanc}, \citenamefont {Antipov}, \citenamefont {Becca}, \citenamefont
  {Bulik}, \citenamefont {Chan}, \citenamefont {Chung}, \citenamefont {Deng},
  \citenamefont {Ferrero}, \citenamefont {Henderson}, \citenamefont
  {Jim{\'e}nez-Hoyos}, \citenamefont {Kozik}, \citenamefont {Liu},
  \citenamefont {Millis}, \citenamefont {Prokof{\textquoteright}ev},
  \citenamefont {Qin}, \citenamefont {Scuseria}, \citenamefont {Shi},
  \citenamefont {Svistunov}, \citenamefont {Tocchio}, \citenamefont {Tupitsyn},
  \citenamefont {White}, \citenamefont {Zhang}, \citenamefont {Zheng},
  \citenamefont {Zhu}, \citenamefont {Gull},\ and\ \citenamefont {{Simons
  Collaboration on the Many-Electron Problem}}}]{LeBlanc:2015ha}%
  \BibitemOpen
  \bibfield  {author} {\bibinfo {author} {\bibfnamefont {J~P~F}\ \bibnamefont
  {LeBlanc}}, \bibinfo {author} {\bibfnamefont {Andrey~E}\ \bibnamefont
  {Antipov}}, \bibinfo {author} {\bibfnamefont {Federico}\ \bibnamefont
  {Becca}}, \bibinfo {author} {\bibfnamefont {Ireneusz~W}\ \bibnamefont
  {Bulik}}, \bibinfo {author} {\bibfnamefont {Garnet Kin-Lic}\ \bibnamefont
  {Chan}}, \bibinfo {author} {\bibfnamefont {Chia-Min}\ \bibnamefont {Chung}},
  \bibinfo {author} {\bibfnamefont {Youjin}\ \bibnamefont {Deng}}, \bibinfo
  {author} {\bibfnamefont {Michel}\ \bibnamefont {Ferrero}}, \bibinfo {author}
  {\bibfnamefont {Thomas~M}\ \bibnamefont {Henderson}}, \bibinfo {author}
  {\bibfnamefont {Carlos~A}\ \bibnamefont {Jim{\'e}nez-Hoyos}}, \bibinfo
  {author} {\bibfnamefont {E}~\bibnamefont {Kozik}}, \bibinfo {author}
  {\bibfnamefont {Xuan-Wen}\ \bibnamefont {Liu}}, \bibinfo {author}
  {\bibfnamefont {Andrew~J}\ \bibnamefont {Millis}}, \bibinfo {author}
  {\bibfnamefont {N~V}\ \bibnamefont {Prokof{\textquoteright}ev}}, \bibinfo
  {author} {\bibfnamefont {Mingpu}\ \bibnamefont {Qin}}, \bibinfo {author}
  {\bibfnamefont {Gustavo~E}\ \bibnamefont {Scuseria}}, \bibinfo {author}
  {\bibfnamefont {Hao}\ \bibnamefont {Shi}}, \bibinfo {author} {\bibfnamefont
  {B~V}\ \bibnamefont {Svistunov}}, \bibinfo {author} {\bibfnamefont {Luca~F}\
  \bibnamefont {Tocchio}}, \bibinfo {author} {\bibfnamefont {I~S}\ \bibnamefont
  {Tupitsyn}}, \bibinfo {author} {\bibfnamefont {Steven~R}\ \bibnamefont
  {White}}, \bibinfo {author} {\bibfnamefont {Shiwei}\ \bibnamefont {Zhang}},
  \bibinfo {author} {\bibfnamefont {Bo-Xiao}\ \bibnamefont {Zheng}}, \bibinfo
  {author} {\bibfnamefont {Zhenyue}\ \bibnamefont {Zhu}}, \bibinfo {author}
  {\bibfnamefont {Emanuel}\ \bibnamefont {Gull}}, \ and\ \bibinfo {author}
  {\bibnamefont {{Simons Collaboration on the Many-Electron Problem}}},\
  }\bibfield  {title} {\enquote {\bibinfo {title} {{Solutions of the
  Two-Dimensional Hubbard Model: Benchmarks and Results from a Wide Range of
  Numerical Algorithms}},}\ }\href {\doibase 10.1103/PhysRevX.5.041041}
  {\bibfield  {journal} {\bibinfo  {journal} {Physical Review X}\ }\textbf
  {\bibinfo {volume} {5}},\ \bibinfo {pages} {041041} (\bibinfo {year}
  {2015})}\BibitemShut {NoStop}%
\bibitem [{\citenamefont {Bloch}\ \emph {et~al.}(2008)\citenamefont {Bloch},
  \citenamefont {Dalibard},\ and\ \citenamefont {Zwerger}}]{Bloch:2008gla}%
  \BibitemOpen
  \bibfield  {author} {\bibinfo {author} {\bibfnamefont {Immanuel}\
  \bibnamefont {Bloch}}, \bibinfo {author} {\bibfnamefont {Jean}\ \bibnamefont
  {Dalibard}}, \ and\ \bibinfo {author} {\bibfnamefont {Wilhelm}\ \bibnamefont
  {Zwerger}},\ }\bibfield  {title} {\enquote {\bibinfo {title} {Many-body
  physics with ultracold gases},}\ }\href {\doibase 10.1103/RevModPhys.80.885}
  {\bibfield  {journal} {\bibinfo  {journal} {Rev. Mod. Phys.}\ }\textbf
  {\bibinfo {volume} {80}},\ \bibinfo {pages} {885--964} (\bibinfo {year}
  {2008})}\BibitemShut {NoStop}%
\bibitem [{\citenamefont {Esslinger}(2010)}]{Esslinger:2010ex}%
  \BibitemOpen
  \bibfield  {author} {\bibinfo {author} {\bibfnamefont {Tilman}\ \bibnamefont
  {Esslinger}},\ }\bibfield  {title} {\enquote {\bibinfo {title}
  {{Fermi-Hubbard Physics with Atoms in an Optical Lattice}},}\ }\href
  {\doibase 10.1146/annurev-conmatphys-070909-104059} {\bibfield  {journal}
  {\bibinfo  {journal} {Annual Review of Condensed Matter Physics}\ }\textbf
  {\bibinfo {volume} {1}},\ \bibinfo {pages} {129--152} (\bibinfo {year}
  {2010})}\BibitemShut {NoStop}%
\bibitem [{\citenamefont {Jordens}\ \emph {et~al.}(2008)\citenamefont
  {Jordens}, \citenamefont {Strohmaier}, \citenamefont {Gunter}, \citenamefont
  {Moritz},\ and\ \citenamefont {Esslinger}}]{Jordens2008}%
  \BibitemOpen
  \bibfield  {author} {\bibinfo {author} {\bibfnamefont {Robert}\ \bibnamefont
  {Jordens}}, \bibinfo {author} {\bibfnamefont {Niels}\ \bibnamefont
  {Strohmaier}}, \bibinfo {author} {\bibfnamefont {Kenneth}\ \bibnamefont
  {Gunter}}, \bibinfo {author} {\bibfnamefont {Henning}\ \bibnamefont
  {Moritz}}, \ and\ \bibinfo {author} {\bibfnamefont {Tilman}\ \bibnamefont
  {Esslinger}},\ }\bibfield  {title} {\enquote {\bibinfo {title} {A mott
  insulator of fermionic atoms in an optical lattice},}\ }\href
  {http://dx.doi.org/10.1038/nature07244} {\bibfield  {journal} {\bibinfo
  {journal} {Nature}\ }\textbf {\bibinfo {volume} {455}},\ \bibinfo {pages}
  {204--207} (\bibinfo {year} {2008})}\BibitemShut {NoStop}%
\bibitem [{\citenamefont {Schneider}\ \emph {et~al.}(2008)\citenamefont
  {Schneider}, \citenamefont {Hackerm{\"u}ller}, \citenamefont {Will},
  \citenamefont {Best}, \citenamefont {Bloch}, \citenamefont {Costi},
  \citenamefont {Helmes}, \citenamefont {Rasch},\ and\ \citenamefont
  {Rosch}}]{Schneider2008}%
  \BibitemOpen
  \bibfield  {author} {\bibinfo {author} {\bibfnamefont {U.}~\bibnamefont
  {Schneider}}, \bibinfo {author} {\bibfnamefont {L.}~\bibnamefont
  {Hackerm{\"u}ller}}, \bibinfo {author} {\bibfnamefont {S.}~\bibnamefont
  {Will}}, \bibinfo {author} {\bibfnamefont {Th.}\ \bibnamefont {Best}},
  \bibinfo {author} {\bibfnamefont {I.}~\bibnamefont {Bloch}}, \bibinfo
  {author} {\bibfnamefont {T.~A.}\ \bibnamefont {Costi}}, \bibinfo {author}
  {\bibfnamefont {R.~W.}\ \bibnamefont {Helmes}}, \bibinfo {author}
  {\bibfnamefont {D.}~\bibnamefont {Rasch}}, \ and\ \bibinfo {author}
  {\bibfnamefont {A.}~\bibnamefont {Rosch}},\ }\bibfield  {title} {\enquote
  {\bibinfo {title} {Metallic and insulating phases of repulsively interacting
  fermions in a 3d optical lattice},}\ }\href {\doibase
  10.1126/science.1165449} {\bibfield  {journal} {\bibinfo  {journal}
  {Science}\ }\textbf {\bibinfo {volume} {322}},\ \bibinfo {pages} {1520--1525}
  (\bibinfo {year} {2008})},\ \Eprint
  {http://arxiv.org/abs/http://science.sciencemag.org/content/322/5907/1520.full.pdf}
  {http://science.sciencemag.org/content/322/5907/1520.full.pdf} \BibitemShut
  {NoStop}%
\bibitem [{\citenamefont {Boll}\ \emph {et~al.}(2016)\citenamefont {Boll},
  \citenamefont {Hilker}, \citenamefont {Salomon}, \citenamefont {Omran},
  \citenamefont {Nespolo}, \citenamefont {Pollet}, \citenamefont {Bloch},\ and\
  \citenamefont {Gross}}]{Boll2016}%
  \BibitemOpen
  \bibfield  {author} {\bibinfo {author} {\bibfnamefont {Martin}\ \bibnamefont
  {Boll}}, \bibinfo {author} {\bibfnamefont {Timon~A.}\ \bibnamefont {Hilker}},
  \bibinfo {author} {\bibfnamefont {Guillaume}\ \bibnamefont {Salomon}},
  \bibinfo {author} {\bibfnamefont {Ahmed}\ \bibnamefont {Omran}}, \bibinfo
  {author} {\bibfnamefont {Jacopo}\ \bibnamefont {Nespolo}}, \bibinfo {author}
  {\bibfnamefont {Lode}\ \bibnamefont {Pollet}}, \bibinfo {author}
  {\bibfnamefont {Immanuel}\ \bibnamefont {Bloch}}, \ and\ \bibinfo {author}
  {\bibfnamefont {Christian}\ \bibnamefont {Gross}},\ }\bibfield  {title}
  {\enquote {\bibinfo {title} {Spin- and density-resolved microscopy of
  antiferromagnetic correlations in fermi-hubbard chains},}\ }\href {\doibase
  10.1126/science.aag1635} {\bibfield  {journal} {\bibinfo  {journal}
  {Science}\ }\textbf {\bibinfo {volume} {353}},\ \bibinfo {pages} {1257--1260}
  (\bibinfo {year} {2016})},\ \Eprint
  {http://arxiv.org/abs/http://science.sciencemag.org/content/353/6305/1257.full.pdf}
  {http://science.sciencemag.org/content/353/6305/1257.full.pdf} \BibitemShut
  {NoStop}%
\bibitem [{\citenamefont {Parsons}\ \emph {et~al.}(2016)\citenamefont
  {Parsons}, \citenamefont {Mazurenko}, \citenamefont {Chiu}, \citenamefont
  {Ji}, \citenamefont {Greif},\ and\ \citenamefont {Greiner}}]{Parsons2016}%
  \BibitemOpen
  \bibfield  {author} {\bibinfo {author} {\bibfnamefont {Maxwell~F.}\
  \bibnamefont {Parsons}}, \bibinfo {author} {\bibfnamefont {Anton}\
  \bibnamefont {Mazurenko}}, \bibinfo {author} {\bibfnamefont {Christie~S.}\
  \bibnamefont {Chiu}}, \bibinfo {author} {\bibfnamefont {Geoffrey}\
  \bibnamefont {Ji}}, \bibinfo {author} {\bibfnamefont {Daniel}\ \bibnamefont
  {Greif}}, \ and\ \bibinfo {author} {\bibfnamefont {Markus}\ \bibnamefont
  {Greiner}},\ }\bibfield  {title} {\enquote {\bibinfo {title} {Site-resolved
  measurement of the spin-correlation function in the fermi-hubbard model},}\
  }\href {\doibase 10.1126/science.aag1430} {\bibfield  {journal} {\bibinfo
  {journal} {Science}\ }\textbf {\bibinfo {volume} {353}},\ \bibinfo {pages}
  {1253--1256} (\bibinfo {year} {2016})},\ \Eprint
  {http://arxiv.org/abs/http://science.sciencemag.org/content/353/6305/1253.full.pdf}
  {http://science.sciencemag.org/content/353/6305/1253.full.pdf} \BibitemShut
  {NoStop}%
\bibitem [{\citenamefont {Cheuk}\ \emph {et~al.}(2016)\citenamefont {Cheuk},
  \citenamefont {Nichols}, \citenamefont {Lawrence}, \citenamefont {Okan},
  \citenamefont {Zhang}, \citenamefont {Khatami}, \citenamefont {Trivedi},
  \citenamefont {Paiva}, \citenamefont {Rigol},\ and\ \citenamefont
  {Zwierlein}}]{Cheuk2016}%
  \BibitemOpen
  \bibfield  {author} {\bibinfo {author} {\bibfnamefont {Lawrence~W.}\
  \bibnamefont {Cheuk}}, \bibinfo {author} {\bibfnamefont {Matthew~A.}\
  \bibnamefont {Nichols}}, \bibinfo {author} {\bibfnamefont {Katherine~R.}\
  \bibnamefont {Lawrence}}, \bibinfo {author} {\bibfnamefont {Melih}\
  \bibnamefont {Okan}}, \bibinfo {author} {\bibfnamefont {Hao}\ \bibnamefont
  {Zhang}}, \bibinfo {author} {\bibfnamefont {Ehsan}\ \bibnamefont {Khatami}},
  \bibinfo {author} {\bibfnamefont {Nandini}\ \bibnamefont {Trivedi}}, \bibinfo
  {author} {\bibfnamefont {Thereza}\ \bibnamefont {Paiva}}, \bibinfo {author}
  {\bibfnamefont {Marcos}\ \bibnamefont {Rigol}}, \ and\ \bibinfo {author}
  {\bibfnamefont {Martin~W.}\ \bibnamefont {Zwierlein}},\ }\bibfield  {title}
  {\enquote {\bibinfo {title} {Observation of spatial charge and spin
  correlations in the 2d fermi-hubbard model},}\ }\href {\doibase
  10.1126/science.aag3349} {\bibfield  {journal} {\bibinfo  {journal}
  {Science}\ }\textbf {\bibinfo {volume} {353}},\ \bibinfo {pages} {1260--1264}
  (\bibinfo {year} {2016})},\ \Eprint
  {http://arxiv.org/abs/http://science.sciencemag.org/content/353/6305/1260.full.pdf}
  {http://science.sciencemag.org/content/353/6305/1260.full.pdf} \BibitemShut
  {NoStop}%
\bibitem [{\citenamefont {{Mazurenko}}\ \emph {et~al.}(2016)\citenamefont
  {{Mazurenko}}, \citenamefont {{Chiu}}, \citenamefont {{Ji}}, \citenamefont
  {{Parsons}}, \citenamefont {{Kan{\'a}sz-Nagy}}, \citenamefont {{Schmidt}},
  \citenamefont {{Grusdt}}, \citenamefont {{Demler}}, \citenamefont {{Greif}},\
  and\ \citenamefont {{Greiner}}}]{Mazurenko2016}%
  \BibitemOpen
  \bibfield  {author} {\bibinfo {author} {\bibfnamefont {A.}~\bibnamefont
  {{Mazurenko}}}, \bibinfo {author} {\bibfnamefont {C.~S.}\ \bibnamefont
  {{Chiu}}}, \bibinfo {author} {\bibfnamefont {G.}~\bibnamefont {{Ji}}},
  \bibinfo {author} {\bibfnamefont {M.~F.}\ \bibnamefont {{Parsons}}}, \bibinfo
  {author} {\bibfnamefont {M.}~\bibnamefont {{Kan{\'a}sz-Nagy}}}, \bibinfo
  {author} {\bibfnamefont {R.}~\bibnamefont {{Schmidt}}}, \bibinfo {author}
  {\bibfnamefont {F.}~\bibnamefont {{Grusdt}}}, \bibinfo {author}
  {\bibfnamefont {E.}~\bibnamefont {{Demler}}}, \bibinfo {author}
  {\bibfnamefont {D.}~\bibnamefont {{Greif}}}, \ and\ \bibinfo {author}
  {\bibfnamefont {M.}~\bibnamefont {{Greiner}}},\ }\bibfield  {title} {\enquote
  {\bibinfo {title} {{Experimental realization of a long-range antiferromagnet
  in the Hubbard model with ultracold atoms}},}\ }\href@noop {} {\bibfield
  {journal} {\bibinfo  {journal} {ArXiv e-prints}\ } (\bibinfo {year}
  {2016})},\ \Eprint {http://arxiv.org/abs/1612.08436} {arXiv:1612.08436
  [cond-mat.quant-gas]} \BibitemShut {NoStop}%
\bibitem [{\citenamefont {Wu}(2008)}]{Wu:2008ea}%
  \BibitemOpen
  \bibfield  {author} {\bibinfo {author} {\bibfnamefont {Congjun}\ \bibnamefont
  {Wu}},\ }\bibfield  {title} {\enquote {\bibinfo {title} {{Orbital Ordering
  and Frustration of p-Band Mott Insulators}},}\ }\href
  {http://link.aps.org/doi/10.1103/PhysRevLett.100.200406} {\bibfield
  {journal} {\bibinfo  {journal} {Physical Review Letters}\ }\textbf {\bibinfo
  {volume} {100}},\ \bibinfo {pages} {200406} (\bibinfo {year}
  {2008})}\BibitemShut {NoStop}%
\bibitem [{\citenamefont {Zhao}\ and\ \citenamefont {Liu}(2008)}]{Zhao:2008ev}%
  \BibitemOpen
  \bibfield  {author} {\bibinfo {author} {\bibfnamefont {Erhai}\ \bibnamefont
  {Zhao}}\ and\ \bibinfo {author} {\bibfnamefont {W}~\bibnamefont {Liu}},\
  }\bibfield  {title} {\enquote {\bibinfo {title} {{Orbital Order in Mott
  Insulators of Spinless p-Band Fermions}},}\ }\href
  {http://link.aps.org/doi/10.1103/PhysRevLett.100.160403} {\bibfield
  {journal} {\bibinfo  {journal} {Physical Review Letters}\ }\textbf {\bibinfo
  {volume} {100}},\ \bibinfo {pages} {160403} (\bibinfo {year}
  {2008})}\BibitemShut {NoStop}%
\bibitem [{\citenamefont {Feiguin}\ and\ \citenamefont
  {Fisher}(2009)}]{Feiguin:2009iu}%
  \BibitemOpen
  \bibfield  {author} {\bibinfo {author} {\bibfnamefont {Adrian~E}\
  \bibnamefont {Feiguin}}\ and\ \bibinfo {author} {\bibfnamefont {Matthew P~A}\
  \bibnamefont {Fisher}},\ }\bibfield  {title} {\enquote {\bibinfo {title}
  {{Exotic Paired States with Anisotropic Spin-Dependent Fermi Surfaces}},}\
  }\href {http://link.aps.org/doi/10.1103/PhysRevLett.103.025303} {\bibfield
  {journal} {\bibinfo  {journal} {Physical Review Letters}\ }\textbf {\bibinfo
  {volume} {103}},\ \bibinfo {pages} {025303} (\bibinfo {year}
  {2009})}\BibitemShut {NoStop}%
\bibitem [{\citenamefont {Feiguin}\ and\ \citenamefont
  {Fisher}(2011)}]{Feiguin:2011jz}%
  \BibitemOpen
  \bibfield  {author} {\bibinfo {author} {\bibfnamefont {Adrian~E}\
  \bibnamefont {Feiguin}}\ and\ \bibinfo {author} {\bibfnamefont {Matthew P~A}\
  \bibnamefont {Fisher}},\ }\bibfield  {title} {\enquote {\bibinfo {title}
  {{Exotic paired phases in ladders with spin-dependent hopping}},}\ }\href
  {http://link.aps.org/doi/10.1103/PhysRevB.83.115104} {\bibfield  {journal}
  {\bibinfo  {journal} {Physical Review B}\ }\textbf {\bibinfo {volume} {83}},\
  \bibinfo {pages} {115104} (\bibinfo {year} {2011})}\BibitemShut {NoStop}%
\bibitem [{\citenamefont {Gukelberger}\ \emph {et~al.}(2014)\citenamefont
  {Gukelberger}, \citenamefont {Kozik}, \citenamefont {Pollet}, \citenamefont
  {Prokof'ev}, \citenamefont {Sigrist}, \citenamefont {Svistunov},\ and\
  \citenamefont {Troyer}}]{Gukelberger:2014dba}%
  \BibitemOpen
  \bibfield  {author} {\bibinfo {author} {\bibfnamefont {Jan}\ \bibnamefont
  {Gukelberger}}, \bibinfo {author} {\bibfnamefont {Evgeny}\ \bibnamefont
  {Kozik}}, \bibinfo {author} {\bibfnamefont {Lode}\ \bibnamefont {Pollet}},
  \bibinfo {author} {\bibfnamefont {Nikolay}\ \bibnamefont {Prokof'ev}},
  \bibinfo {author} {\bibfnamefont {Manfred}\ \bibnamefont {Sigrist}}, \bibinfo
  {author} {\bibfnamefont {Boris}\ \bibnamefont {Svistunov}}, \ and\ \bibinfo
  {author} {\bibfnamefont {Matthias}\ \bibnamefont {Troyer}},\ }\bibfield
  {title} {\enquote {\bibinfo {title} {{p-Wave Superfluidity by Spin-Nematic
  Fermi Surface Deformation}},}\ }\href
  {http://link.aps.org/doi/10.1103/PhysRevLett.113.195301} {\bibfield
  {journal} {\bibinfo  {journal} {Physical Review Letters}\ }\textbf {\bibinfo
  {volume} {113}},\ \bibinfo {pages} {195301} (\bibinfo {year}
  {2014})}\BibitemShut {NoStop}%
\bibitem [{\citenamefont {Xu}\ \emph {et~al.}(2015)\citenamefont {Xu},
  \citenamefont {Li},\ and\ \citenamefont {Wu}}]{Xu:2015gx}%
  \BibitemOpen
  \bibfield  {author} {\bibinfo {author} {\bibfnamefont {Shenglong}\
  \bibnamefont {Xu}}, \bibinfo {author} {\bibfnamefont {Yi}~\bibnamefont {Li}},
  \ and\ \bibinfo {author} {\bibfnamefont {Congjun}\ \bibnamefont {Wu}},\
  }\bibfield  {title} {\enquote {\bibinfo {title} {{Sign-Problem-Free Quantum
  Monte Carlo Study on Thermodynamic Properties and Magnetic Phase Transitions
  in Orbital-Active Itinerant Ferromagnets}},}\ }\href
  {http://link.aps.org/doi/10.1103/PhysRevX.5.021032} {\bibfield  {journal}
  {\bibinfo  {journal} {Physical Review X}\ }\textbf {\bibinfo {volume} {5}},\
  \bibinfo {pages} {021032} (\bibinfo {year} {2015})}\BibitemShut {NoStop}%
\bibitem [{\citenamefont {Berg}\ \emph {et~al.}(2012)\citenamefont {Berg},
  \citenamefont {Metlitski},\ and\ \citenamefont {Sachdev}}]{Berg:2012ie}%
  \BibitemOpen
  \bibfield  {author} {\bibinfo {author} {\bibfnamefont {E}~\bibnamefont
  {Berg}}, \bibinfo {author} {\bibfnamefont {M~A}\ \bibnamefont {Metlitski}}, \
  and\ \bibinfo {author} {\bibfnamefont {S}~\bibnamefont {Sachdev}},\
  }\bibfield  {title} {\enquote {\bibinfo {title} {{Sign-Problem-Free Quantum
  Monte Carlo of the Onset of Antiferromagnetism in Metals}},}\ }\href
  {http://www.sciencemag.org/cgi/doi/10.1126/science.1227769} {\bibfield
  {journal} {\bibinfo  {journal} {Science}\ }\textbf {\bibinfo {volume}
  {338}},\ \bibinfo {pages} {1606--1609} (\bibinfo {year} {2012})}\BibitemShut
  {NoStop}%
\bibitem [{\citenamefont {Prokof'ev}\ \emph {et~al.}(1998)\citenamefont
  {Prokof'ev}, \citenamefont {Svistunov},\ and\ \citenamefont
  {Tupitsyn}}]{Prokofev1998}%
  \BibitemOpen
  \bibfield  {author} {\bibinfo {author} {\bibfnamefont {N.~V.}\ \bibnamefont
  {Prokof'ev}}, \bibinfo {author} {\bibfnamefont {B.~V.}\ \bibnamefont
  {Svistunov}}, \ and\ \bibinfo {author} {\bibfnamefont {I.~S.}\ \bibnamefont
  {Tupitsyn}},\ }\bibfield  {title} {\enquote {\bibinfo {title} {Exact,
  complete, and universal continuous-time worldline monte carlo approach to the
  statistics of discrete quantum systems},}\ }\href {\doibase 10.1134/1.558661}
  {\bibfield  {journal} {\bibinfo  {journal} {Journal of Experimental and
  Theoretical Physics}\ }\textbf {\bibinfo {volume} {87}},\ \bibinfo {pages}
  {310--321} (\bibinfo {year} {1998})}\BibitemShut {NoStop}%
\bibitem [{\citenamefont {Pollet}\ \emph {et~al.}(2007)\citenamefont {Pollet},
  \citenamefont {Houcke},\ and\ \citenamefont {Rombouts}}]{Pollet2007}%
  \BibitemOpen
  \bibfield  {author} {\bibinfo {author} {\bibfnamefont {Lode}\ \bibnamefont
  {Pollet}}, \bibinfo {author} {\bibfnamefont {Kris~Van}\ \bibnamefont
  {Houcke}}, \ and\ \bibinfo {author} {\bibfnamefont {Stefan~M.A.}\
  \bibnamefont {Rombouts}},\ }\bibfield  {title} {\enquote {\bibinfo {title}
  {Engineering local optimality in quantum monte carlo algorithms},}\ }\href
  {\doibase http://dx.doi.org/10.1016/j.jcp.2007.03.013} {\bibfield  {journal}
  {\bibinfo  {journal} {Journal of Computational Physics}\ }\textbf {\bibinfo
  {volume} {225}},\ \bibinfo {pages} {2249 -- 2266} (\bibinfo {year}
  {2007})}\BibitemShut {NoStop}%
\bibitem [{\citenamefont {{Van Houcke}}\ \emph {et~al.}(2010)\citenamefont
  {{Van Houcke}}, \citenamefont {Kozik}, \citenamefont {Prokof'ev},\ and\
  \citenamefont {Svistunov}}]{VanHoucke2010}%
  \BibitemOpen
  \bibfield  {author} {\bibinfo {author} {\bibfnamefont {Kris}\ \bibnamefont
  {{Van Houcke}}}, \bibinfo {author} {\bibfnamefont {Evgeny}\ \bibnamefont
  {Kozik}}, \bibinfo {author} {\bibfnamefont {N.}~\bibnamefont {Prokof'ev}}, \
  and\ \bibinfo {author} {\bibfnamefont {B.}~\bibnamefont {Svistunov}},\
  }\bibfield  {title} {\enquote {\bibinfo {title} {{Diagrammatic Monte
  Carlo}},}\ }\href {\doibase 10.1016/j.phpro.2010.09.034} {\bibfield
  {journal} {\bibinfo  {journal} {Phys. Procedia}\ }\textbf {\bibinfo {volume}
  {6}},\ \bibinfo {pages} {95--105} (\bibinfo {year} {2010})},\ \Eprint
  {http://arxiv.org/abs/0802.2923} {arXiv:0802.2923} \BibitemShut {NoStop}%
\bibitem [{\citenamefont {Gukelberger}(2015)}]{gukelberger2015diss}%
  \BibitemOpen
  \bibfield  {author} {\bibinfo {author} {\bibfnamefont {Jan}\ \bibnamefont
  {Gukelberger}},\ }\emph {\bibinfo {title} {{From non-unitary anyons to
  unconventional superfluidity}}},\ \href {\doibase 10.3929/ethz-a-010451939}
  {Ph.D. thesis},\ \bibinfo  {school} {ETH Zurich} (\bibinfo {year} {2015}),\
  \bibinfo {note} {{DOI:0.3929/ethz-a-010451939}}\BibitemShut {NoStop}%
\bibitem [{\citenamefont {Iazzi}\ and\ \citenamefont
  {Troyer}(2015)}]{Iazzi:2015hi}%
  \BibitemOpen
  \bibfield  {author} {\bibinfo {author} {\bibfnamefont {Mauro}\ \bibnamefont
  {Iazzi}}\ and\ \bibinfo {author} {\bibfnamefont {Matthias}\ \bibnamefont
  {Troyer}},\ }\bibfield  {title} {\enquote {\bibinfo {title} {{Efficient
  continuous-time quantum Monte Carlo algorithm for fermionic lattice
  models}},}\ }\href {http://link.aps.org/doi/10.1103/PhysRevB.91.241118}
  {\bibfield  {journal} {\bibinfo  {journal} {Physical Review B}\ }\textbf
  {\bibinfo {volume} {91}},\ \bibinfo {pages} {241118} (\bibinfo {year}
  {2015})}\BibitemShut {NoStop}%
\bibitem [{\citenamefont {Wang}\ \emph
  {et~al.}(2015{\natexlab{a}})\citenamefont {Wang}, \citenamefont {Iazzi},
  \citenamefont {Corboz},\ and\ \citenamefont {Troyer}}]{2015PhRvB..91w5151W}%
  \BibitemOpen
  \bibfield  {author} {\bibinfo {author} {\bibfnamefont {Lei}\ \bibnamefont
  {Wang}}, \bibinfo {author} {\bibfnamefont {Mauro}\ \bibnamefont {Iazzi}},
  \bibinfo {author} {\bibfnamefont {Philippe}\ \bibnamefont {Corboz}}, \ and\
  \bibinfo {author} {\bibfnamefont {Matthias}\ \bibnamefont {Troyer}},\
  }\bibfield  {title} {\enquote {\bibinfo {title} {{Efficient continuous-time
  quantum Monte Carlo method for the ground state of correlated fermions}},}\
  }\href
  {http://adsabs.harvard.edu/cgi-bin/nph-data_query?bibcode=2015PhRvB..91w5151W&link_type=ABSTRACT}
  {\bibfield  {journal} {\bibinfo  {journal} {Physical Review B}\ }\textbf
  {\bibinfo {volume} {91}},\ \bibinfo {pages} {235151} (\bibinfo {year}
  {2015}{\natexlab{a}})}\BibitemShut {NoStop}%
\bibitem [{\citenamefont {Cazalilla}\ \emph {et~al.}(2011)\citenamefont
  {Cazalilla}, \citenamefont {Citro}, \citenamefont {Giamarchi}, \citenamefont
  {Orignac},\ and\ \citenamefont {Rigol}}]{Cazalilla2011}%
  \BibitemOpen
  \bibfield  {author} {\bibinfo {author} {\bibfnamefont {M.~A.}\ \bibnamefont
  {Cazalilla}}, \bibinfo {author} {\bibfnamefont {R.}~\bibnamefont {Citro}},
  \bibinfo {author} {\bibfnamefont {T.}~\bibnamefont {Giamarchi}}, \bibinfo
  {author} {\bibfnamefont {E.}~\bibnamefont {Orignac}}, \ and\ \bibinfo
  {author} {\bibfnamefont {M.}~\bibnamefont {Rigol}},\ }\bibfield  {title}
  {\enquote {\bibinfo {title} {{One dimensional bosons: From condensed matter
  systems to ultracold gases}},}\ }\href {\doibase 10.1103/RevModPhys.83.1405}
  {\bibfield  {journal} {\bibinfo  {journal} {Reviews of Modern Physics}\
  }\textbf {\bibinfo {volume} {83}},\ \bibinfo {pages} {1405--1466} (\bibinfo
  {year} {2011})},\ \Eprint {http://arxiv.org/abs/1101.5337} {arXiv:1101.5337}
  \BibitemShut {NoStop}%
\bibitem [{\citenamefont {Giamarchi}(2003)}]{Giamarchi_book}%
  \BibitemOpen
  \bibfield  {author} {\bibinfo {author} {\bibfnamefont {T.}~\bibnamefont
  {Giamarchi}},\ }\href {https://books.google.de/books?id=GVeuKZLGMZ0C} {\emph
  {\bibinfo {title} {Quantum Physics in One Dimension}}},\ International Series
  of Monographs on Physics\ (\bibinfo  {publisher} {Clarendon Press},\ \bibinfo
  {year} {2003})\BibitemShut {NoStop}%
\bibitem [{\citenamefont {Prokof'ev}\ and\ \citenamefont
  {Svistunov}(2007)}]{prokofev2007bdm}%
  \BibitemOpen
  \bibfield  {author} {\bibinfo {author} {\bibfnamefont {Nikolay}\ \bibnamefont
  {Prokof'ev}}\ and\ \bibinfo {author} {\bibfnamefont {Boris}\ \bibnamefont
  {Svistunov}},\ }\bibfield  {title} {\enquote {\bibinfo {title} {{Bold
  diagrammatic Monte Carlo technique: When the sign problem is welcome}},}\
  }\href {\doibase 10.1103/PhysRevLett.99.250201} {\bibfield  {journal}
  {\bibinfo  {journal} {Phys. Rev. Lett.}\ }\textbf {\bibinfo {volume} {99}},\
  \bibinfo {pages} {250201} (\bibinfo {year} {2007})}\BibitemShut {NoStop}%
\bibitem [{\citenamefont {Prokof'ev}\ and\ \citenamefont
  {Svistunov}(2008)}]{prokofev2008fpp}%
  \BibitemOpen
  \bibfield  {author} {\bibinfo {author} {\bibfnamefont {Nikolay}\ \bibnamefont
  {Prokof'ev}}\ and\ \bibinfo {author} {\bibfnamefont {Boris}\ \bibnamefont
  {Svistunov}},\ }\bibfield  {title} {\enquote {\bibinfo {title}
  {{Fermi-polaron problem: Diagrammatic Monte Carlo method for divergent
  sign-alternating series}},}\ }\href {\doibase 10.1103/PhysRevB.77.020408}
  {\bibfield  {journal} {\bibinfo  {journal} {Phys. Rev. B}\ }\textbf {\bibinfo
  {volume} {77}},\ \bibinfo {pages} {020408} (\bibinfo {year}
  {2008})}\BibitemShut {NoStop}%
\bibitem [{\citenamefont {Huffman}\ and\ \citenamefont
  {Chandrasekharan}(2014)}]{Huffman:2014fj}%
  \BibitemOpen
  \bibfield  {author} {\bibinfo {author} {\bibfnamefont {Emilie~Fulton}\
  \bibnamefont {Huffman}}\ and\ \bibinfo {author} {\bibfnamefont {Shailesh}\
  \bibnamefont {Chandrasekharan}},\ }\bibfield  {title} {\enquote {\bibinfo
  {title} {{Solution to sign problems in half-filled spin-polarized electronic
  systems}},}\ }\href {http://link.aps.org/doi/10.1103/PhysRevB.89.111101}
  {\bibfield  {journal} {\bibinfo  {journal} {Physical Review B}\ }\textbf
  {\bibinfo {volume} {89}},\ \bibinfo {pages} {111101} (\bibinfo {year}
  {2014})}\BibitemShut {NoStop}%
\bibitem [{\citenamefont {Li}\ \emph {et~al.}(2015)\citenamefont {Li},
  \citenamefont {Jiang},\ and\ \citenamefont {Yao}}]{Li:2015jf}%
  \BibitemOpen
  \bibfield  {author} {\bibinfo {author} {\bibfnamefont {Zi-Xiang}\
  \bibnamefont {Li}}, \bibinfo {author} {\bibfnamefont {Yi-Fan}\ \bibnamefont
  {Jiang}}, \ and\ \bibinfo {author} {\bibfnamefont {Hong}\ \bibnamefont
  {Yao}},\ }\bibfield  {title} {\enquote {\bibinfo {title} {{Solving the
  fermion sign problem in quantum Monte Carlo simulations by Majorana
  representation}},}\ }\href
  {http://link.aps.org/doi/10.1103/PhysRevB.91.241117} {\bibfield  {journal}
  {\bibinfo  {journal} {Physical Review B}\ }\textbf {\bibinfo {volume} {91}},\
  \bibinfo {pages} {241117} (\bibinfo {year} {2015})}\BibitemShut {NoStop}%
\bibitem [{\citenamefont {Wang}\ \emph
  {et~al.}(2015{\natexlab{b}})\citenamefont {Wang}, \citenamefont {Liu},
  \citenamefont {Iazzi}, \citenamefont {Troyer},\ and\ \citenamefont
  {Harcos}}]{Wang:2015hm}%
  \BibitemOpen
  \bibfield  {author} {\bibinfo {author} {\bibfnamefont {Lei}\ \bibnamefont
  {Wang}}, \bibinfo {author} {\bibfnamefont {Ye-Hua}\ \bibnamefont {Liu}},
  \bibinfo {author} {\bibfnamefont {Mauro}\ \bibnamefont {Iazzi}}, \bibinfo
  {author} {\bibfnamefont {Matthias}\ \bibnamefont {Troyer}}, \ and\ \bibinfo
  {author} {\bibfnamefont {Gergely}\ \bibnamefont {Harcos}},\ }\bibfield
  {title} {\enquote {\bibinfo {title} {{Split Orthogonal Group: A Guiding
  Principle for Sign-Problem-Free Fermionic Simulations}},}\ }\href {\doibase
  10.1103/PhysRevLett.115.250601} {\bibfield  {journal} {\bibinfo  {journal}
  {Physical Review Letters}\ }\textbf {\bibinfo {volume} {115}},\ \bibinfo
  {pages} {250601} (\bibinfo {year} {2015}{\natexlab{b}})}\BibitemShut
  {NoStop}%
\bibitem [{\citenamefont {Wei}\ \emph {et~al.}(2016)\citenamefont {Wei},
  \citenamefont {Wu}, \citenamefont {Li}, \citenamefont {Zhang},\ and\
  \citenamefont {Xiang}}]{PhysRevLett.116.250601}%
  \BibitemOpen
  \bibfield  {author} {\bibinfo {author} {\bibfnamefont {Z.~C.}\ \bibnamefont
  {Wei}}, \bibinfo {author} {\bibfnamefont {Congjun}\ \bibnamefont {Wu}},
  \bibinfo {author} {\bibfnamefont {Yi}~\bibnamefont {Li}}, \bibinfo {author}
  {\bibfnamefont {Shiwei}\ \bibnamefont {Zhang}}, \ and\ \bibinfo {author}
  {\bibfnamefont {T.}~\bibnamefont {Xiang}},\ }\bibfield  {title} {\enquote
  {\bibinfo {title} {Majorana positivity and the fermion sign problem of
  quantum monte carlo simulations},}\ }\href {\doibase
  10.1103/PhysRevLett.116.250601} {\bibfield  {journal} {\bibinfo  {journal}
  {Phys. Rev. Lett.}\ }\textbf {\bibinfo {volume} {116}},\ \bibinfo {pages}
  {250601} (\bibinfo {year} {2016})}\BibitemShut {NoStop}%
\bibitem [{\citenamefont {Liu}\ and\ \citenamefont {Wang}(2015)}]{Liu:2015kx}%
  \BibitemOpen
  \bibfield  {author} {\bibinfo {author} {\bibfnamefont {Ye-Hua}\ \bibnamefont
  {Liu}}\ and\ \bibinfo {author} {\bibfnamefont {Lei}\ \bibnamefont {Wang}},\
  }\bibfield  {title} {\enquote {\bibinfo {title} {{Quantum Monte Carlo study
  of mass-imbalanced Hubbard models}},}\ }\href {\doibase
  10.1103/PhysRevB.92.235129} {\bibfield  {journal} {\bibinfo  {journal}
  {Physical Review B}\ }\textbf {\bibinfo {volume} {92}},\ \bibinfo {pages}
  {235129} (\bibinfo {year} {2015})}\BibitemShut {NoStop}%
\bibitem [{\citenamefont {Jr}\ \emph {et~al.}(1985)\citenamefont {Jr},
  \citenamefont {Scalapino},\ and\ \citenamefont {Grant}}]{1402-4896-32-4-016}%
  \BibitemOpen
  \bibfield  {author} {\bibinfo {author} {\bibfnamefont {E~Loh}\ \bibnamefont
  {Jr}}, \bibinfo {author} {\bibfnamefont {D~J}\ \bibnamefont {Scalapino}}, \
  and\ \bibinfo {author} {\bibfnamefont {P~M}\ \bibnamefont {Grant}},\
  }\bibfield  {title} {\enquote {\bibinfo {title} {Monte carlo simulations of
  the quantum xxz model in two dimensions},}\ }\href
  {http://stacks.iop.org/1402-4896/32/i=4/a=016} {\bibfield  {journal}
  {\bibinfo  {journal} {Physica Scripta}\ }\textbf {\bibinfo {volume} {32}},\
  \bibinfo {pages} {327} (\bibinfo {year} {1985})}\BibitemShut {NoStop}%
\bibitem [{\citenamefont {Mermin}\ and\ \citenamefont
  {Wagner}(1966)}]{MerminWagnerTheorem1966}%
  \BibitemOpen
  \bibfield  {author} {\bibinfo {author} {\bibfnamefont {N.~D.}\ \bibnamefont
  {Mermin}}\ and\ \bibinfo {author} {\bibfnamefont {H.}~\bibnamefont
  {Wagner}},\ }\bibfield  {title} {\enquote {\bibinfo {title} {Absence of
  ferromagnetism or antiferromagnetism in one- or two-dimensional isotropic
  heisenberg models},}\ }\href {\doibase 10.1103/PhysRevLett.17.1133}
  {\bibfield  {journal} {\bibinfo  {journal} {Phys. Rev. Lett.}\ }\textbf
  {\bibinfo {volume} {17}},\ \bibinfo {pages} {1133--1136} (\bibinfo {year}
  {1966})}\BibitemShut {NoStop}%
\bibitem [{Note1()}]{Note1}%
  \BibitemOpen
  \bibinfo {note} {Cf.\ Ref.~\cite {Gukelberger:2014dba}. Note that, while
  there is a direct mapping between results for the half-filled attractive and
  repulsive models, the doped repulsive model maps into the attractive model in
  a magnetic field and vice versa~\cite {ho2009qsh}.}\BibitemShut {Stop}%
\bibitem [{\citenamefont {Mandel}\ \emph {et~al.}(2003)\citenamefont {Mandel},
  \citenamefont {Greiner}, \citenamefont {Widera}, \citenamefont {Rom},
  \citenamefont {H\"ansch},\ and\ \citenamefont
  {Bloch}}]{PhysRevLett.91.010407}%
  \BibitemOpen
  \bibfield  {author} {\bibinfo {author} {\bibfnamefont {Olaf}\ \bibnamefont
  {Mandel}}, \bibinfo {author} {\bibfnamefont {Markus}\ \bibnamefont
  {Greiner}}, \bibinfo {author} {\bibfnamefont {Artur}\ \bibnamefont {Widera}},
  \bibinfo {author} {\bibfnamefont {Tim}\ \bibnamefont {Rom}}, \bibinfo
  {author} {\bibfnamefont {Theodor~W.}\ \bibnamefont {H\"ansch}}, \ and\
  \bibinfo {author} {\bibfnamefont {Immanuel}\ \bibnamefont {Bloch}},\
  }\bibfield  {title} {\enquote {\bibinfo {title} {Coherent transport of
  neutral atoms in spin-dependent optical lattice potentials},}\ }\href
  {\doibase 10.1103/PhysRevLett.91.010407} {\bibfield  {journal} {\bibinfo
  {journal} {Phys. Rev. Lett.}\ }\textbf {\bibinfo {volume} {91}},\ \bibinfo
  {pages} {010407} (\bibinfo {year} {2003})}\BibitemShut {NoStop}%
\bibitem [{\citenamefont {M\"uller}\ \emph {et~al.}(2007)\citenamefont
  {M\"uller}, \citenamefont {F\"olling}, \citenamefont {Widera},\ and\
  \citenamefont {Bloch}}]{PhysRevLett.99.200405}%
  \BibitemOpen
  \bibfield  {author} {\bibinfo {author} {\bibfnamefont {Torben}\ \bibnamefont
  {M\"uller}}, \bibinfo {author} {\bibfnamefont {Simon}\ \bibnamefont
  {F\"olling}}, \bibinfo {author} {\bibfnamefont {Artur}\ \bibnamefont
  {Widera}}, \ and\ \bibinfo {author} {\bibfnamefont {Immanuel}\ \bibnamefont
  {Bloch}},\ }\bibfield  {title} {\enquote {\bibinfo {title} {State preparation
  and dynamics of ultracold atoms in higher lattice orbitals},}\ }\href
  {\doibase 10.1103/PhysRevLett.99.200405} {\bibfield  {journal} {\bibinfo
  {journal} {Phys. Rev. Lett.}\ }\textbf {\bibinfo {volume} {99}},\ \bibinfo
  {pages} {200405} (\bibinfo {year} {2007})}\BibitemShut {NoStop}%
\bibitem [{\citenamefont {Wirth}\ \emph {et~al.}(2010)\citenamefont {Wirth},
  \citenamefont {{\"O}lschl{\"a}ger},\ and\ \citenamefont
  {Hemmerich}}]{Wirth:2010bh}%
  \BibitemOpen
  \bibfield  {author} {\bibinfo {author} {\bibfnamefont {Georg}\ \bibnamefont
  {Wirth}}, \bibinfo {author} {\bibfnamefont {Matthias}\ \bibnamefont
  {{\"O}lschl{\"a}ger}}, \ and\ \bibinfo {author} {\bibfnamefont {Andreas}\
  \bibnamefont {Hemmerich}},\ }\bibfield  {title} {\enquote {\bibinfo {title}
  {{Evidence for orbital superfluidity in the P-band of a bipartite optical
  square lattice}},}\ }\href {\doibase 10.1038/nphys1857} {\bibfield  {journal}
  {\bibinfo  {journal} {Nature Physics}\ }\textbf {\bibinfo {volume} {7}},\
  \bibinfo {pages} {147--153} (\bibinfo {year} {2010})}\BibitemShut {NoStop}%
\bibitem [{\citenamefont {Jotzu}\ \emph {et~al.}(2015)\citenamefont {Jotzu},
  \citenamefont {Messer}, \citenamefont {G\"org}, \citenamefont {Greif},
  \citenamefont {Desbuquois},\ and\ \citenamefont
  {Esslinger}}]{PhysRevLett.115.073002}%
  \BibitemOpen
  \bibfield  {author} {\bibinfo {author} {\bibfnamefont {Gregor}\ \bibnamefont
  {Jotzu}}, \bibinfo {author} {\bibfnamefont {Michael}\ \bibnamefont {Messer}},
  \bibinfo {author} {\bibfnamefont {Frederik}\ \bibnamefont {G\"org}}, \bibinfo
  {author} {\bibfnamefont {Daniel}\ \bibnamefont {Greif}}, \bibinfo {author}
  {\bibfnamefont {R\'emi}\ \bibnamefont {Desbuquois}}, \ and\ \bibinfo {author}
  {\bibfnamefont {Tilman}\ \bibnamefont {Esslinger}},\ }\bibfield  {title}
  {\enquote {\bibinfo {title} {Creating state-dependent lattices for ultracold
  fermions by magnetic gradient modulation},}\ }\href {\doibase
  10.1103/PhysRevLett.115.073002} {\bibfield  {journal} {\bibinfo  {journal}
  {Phys. Rev. Lett.}\ }\textbf {\bibinfo {volume} {115}},\ \bibinfo {pages}
  {073002} (\bibinfo {year} {2015})}\BibitemShut {NoStop}%
\bibitem [{\citenamefont {Onsager}(1944)}]{PhysRev.65.117}%
  \BibitemOpen
  \bibfield  {author} {\bibinfo {author} {\bibfnamefont {Lars}\ \bibnamefont
  {Onsager}},\ }\bibfield  {title} {\enquote {\bibinfo {title} {Crystal
  statistics. i. a two-dimensional model with an order-disorder transition},}\
  }\href {\doibase 10.1103/PhysRev.65.117} {\bibfield  {journal} {\bibinfo
  {journal} {Phys. Rev.}\ }\textbf {\bibinfo {volume} {65}},\ \bibinfo {pages}
  {117--149} (\bibinfo {year} {1944})}\BibitemShut {NoStop}%
\bibitem [{\citenamefont {J\"ordens}\ \emph {et~al.}(2010)\citenamefont
  {J\"ordens}, \citenamefont {Tarruell}, \citenamefont {Greif}, \citenamefont
  {Uehlinger}, \citenamefont {Strohmaier}, \citenamefont {Moritz},
  \citenamefont {Esslinger}, \citenamefont {De~Leo}, \citenamefont {Kollath},
  \citenamefont {Georges}, \citenamefont {Scarola}, \citenamefont {Pollet},
  \citenamefont {Burovski}, \citenamefont {Kozik},\ and\ \citenamefont
  {Troyer}}]{PhysRevLett.104.180401}%
  \BibitemOpen
  \bibfield  {author} {\bibinfo {author} {\bibfnamefont {R.}~\bibnamefont
  {J\"ordens}}, \bibinfo {author} {\bibfnamefont {L.}~\bibnamefont {Tarruell}},
  \bibinfo {author} {\bibfnamefont {D.}~\bibnamefont {Greif}}, \bibinfo
  {author} {\bibfnamefont {T.}~\bibnamefont {Uehlinger}}, \bibinfo {author}
  {\bibfnamefont {N.}~\bibnamefont {Strohmaier}}, \bibinfo {author}
  {\bibfnamefont {H.}~\bibnamefont {Moritz}}, \bibinfo {author} {\bibfnamefont
  {T.}~\bibnamefont {Esslinger}}, \bibinfo {author} {\bibfnamefont
  {L.}~\bibnamefont {De~Leo}}, \bibinfo {author} {\bibfnamefont
  {C.}~\bibnamefont {Kollath}}, \bibinfo {author} {\bibfnamefont
  {A.}~\bibnamefont {Georges}}, \bibinfo {author} {\bibfnamefont
  {V.}~\bibnamefont {Scarola}}, \bibinfo {author} {\bibfnamefont
  {L.}~\bibnamefont {Pollet}}, \bibinfo {author} {\bibfnamefont
  {E.}~\bibnamefont {Burovski}}, \bibinfo {author} {\bibfnamefont
  {E.}~\bibnamefont {Kozik}}, \ and\ \bibinfo {author} {\bibfnamefont
  {M.}~\bibnamefont {Troyer}},\ }\bibfield  {title} {\enquote {\bibinfo {title}
  {Quantitative determination of temperature in the approach to magnetic order
  of ultracold fermions in an optical lattice},}\ }\href {\doibase
  10.1103/PhysRevLett.104.180401} {\bibfield  {journal} {\bibinfo  {journal}
  {Phys. Rev. Lett.}\ }\textbf {\bibinfo {volume} {104}},\ \bibinfo {pages}
  {180401} (\bibinfo {year} {2010})}\BibitemShut {NoStop}%
\bibitem [{\citenamefont {{B Bauer \textit{et al.}}}(2011)}]{alps20}%
  \BibitemOpen
  \bibfield  {author} {\bibinfo {author} {\bibnamefont {{B Bauer \textit{et
  al.}}}},\ }\bibfield  {title} {\enquote {\bibinfo {title} {The alps project
  release 2.0: open source software for strongly correlated systems},}\ }\href
  {http://stacks.iop.org/1742-5468/2011/i=05/a=P05001} {\bibfield  {journal}
  {\bibinfo  {journal} {J. Stat. Mech.}\ ,\ \bibinfo {pages} {P05001}}
  (\bibinfo {year} {2011})}\BibitemShut {NoStop}%
\bibitem [{\citenamefont {{A.F. Albuquerque \textit{et al.}}}(2007)}]{alps13}%
  \BibitemOpen
  \bibfield  {author} {\bibinfo {author} {\bibnamefont {{A.F. Albuquerque
  \textit{et al.}}}},\ }\bibfield  {title} {\enquote {\bibinfo {title} {The
  alps project release 1.3: Open-source software for strongly correlated
  systems},}\ }\href {\doibase 10.1016/j.jmmm.2006.10.304} {\bibfield
  {journal} {\bibinfo  {journal} {Journal of Magnetism and Magnetic Materials}\
  }\textbf {\bibinfo {volume} {310}},\ \bibinfo {pages} {1187 -- 1193}
  (\bibinfo {year} {2007})},\ \bibinfo {note} {proceedings of the 17th
  International Conference on Magnetism The International Conference on
  Magnetism}\BibitemShut {NoStop}%
\bibitem [{\citenamefont {Ho}\ \emph {et~al.}(2009)\citenamefont {Ho},
  \citenamefont {Cazalilla},\ and\ \citenamefont {Giamarchi}}]{ho2009qsh}%
  \BibitemOpen
  \bibfield  {author} {\bibinfo {author} {\bibfnamefont {A~F}\ \bibnamefont
  {Ho}}, \bibinfo {author} {\bibfnamefont {M~A}\ \bibnamefont {Cazalilla}}, \
  and\ \bibinfo {author} {\bibfnamefont {T}~\bibnamefont {Giamarchi}},\
  }\bibfield  {title} {\enquote {\bibinfo {title} {{Quantum simulation of the
  Hubbard model: The attractive route}},}\ }\href {\doibase
  10.1103/PhysRevA.79.033620} {\bibfield  {journal} {\bibinfo  {journal}
  {Physical Review A}\ }\textbf {\bibinfo {volume} {79}},\ \bibinfo {pages}
  {033620} (\bibinfo {year} {2009})}\BibitemShut {NoStop}%
\end{thebibliography}%

\end{document}